%% file: paper.tex
\documentstyle[epsf]{mn}

\title{Dynamical evolution of star clusters in tidal fields}
\author[H. Baumgardt]
       {Holger Baumgardt, Junichiro Makino\\
        Department of Astronomy, School of Science, The University of Tokyo,
            7-3-1 Hongo, Bunkyo-ku, Tokyo 113-0033, Japan}
\date{Accepted .
      Received ;
      in original form }

\pagerange{\pageref{firstpage}--\pageref{lastpage}}
\pubyear{2002}

\begin{document}

\maketitle

\label{firstpage}

\begin{abstract}
We report results of a large set of $N$-body calculations aimed to study the evolution of multi-mass star 
clusters in external tidal fields. 
Our clusters start with the same initial mass-functions, but varying particle numbers, orbital types 
and density profiles. Our main focus is to study how the stellar mass-function and other cluster parameters
change under the combined influence of stellar evolution, two-body relaxation and the external tidal field. 

We find that the lifetimes of star clusters moving on similar orbits scale as $T \sim T_{RH}^x$, where 
$T_{RH}$ is the relaxation time, and the exponent $x$ depends on the initial concentration of the
cluster and is around $x \approx 0.75$. The scaling law does not change significantly if one goes from
circular orbits to eccentric ones. From the results for the lifetimes, we predict 
that between 53\% to 67\% of all galactic globular clusters will be destroyed within the next Hubble time. 
Low-mass stars are preferentially lost and the depletion is strong enough to turn initially increasing mass-functions
into mass-functions which decrease towards the low-mass end. The details of this depletion are insensitive  
to the starting condition of the cluster and can be characterised as a function of
a single variable, as e.g. the fraction of time spent until total cluster dissolution. 

The preferential depletion of low-mass stars from star clusters leads to a decrease of their
mass-to-light ratios except for a short period close to final dissolution, when the mass-fraction in
form of compact remnants starts to dominate.
The fraction of compact remnants is increasing throughout the evolution. They are 
more strongly concentrated towards the cluster cores than main-sequence stars and
their mass-fraction in the center can reach 95\% or more around and after core-collapse. 

For a sample of galactic globular clusters with well
observed parameters, we find a correlation between the observed slope of
the mass-function and the lifetimes predicted by us.
It seems possible that galactic globular clusters started with a mass-function similar to what one
observes for the average mass-function of the galactic disc and bulge. 
\end{abstract}

\begin{keywords}
methods: methods: numerical, stellar dynamics - globular clusters: general.
\end{keywords}

\section{Introduction}

The initial mass function (IMF) of stars is an important quantity for astronomy since it influences a 
wide variety of astrophysical processes, like e.g. the chemical evolution of galaxies, the dynamical
evolution of star clusters and the determination of the absolute star formation rate. It is also important 
for theory, since any theory of star formation has to be able to reproduce it.   

One of the best environments to study the mass-function of stars are globular clusters, since each globular
cluster contains a relatively large sample of stars with similar distances, chemical composition and 
ages. In addition, the fraction of binary stars is relatively low in globular clusters, of order 15\% to 
20\% (Albrow et al.\ 2001, 
Elson et al.\ 1998), and is thought to effect the determination 
of the mass-function only in the cluster centers (Rubenstein \& Bailyn 1999). 

Thanks to the availability of the Hubble Space Telescope and large ground-based telescopes, it is nowadays
possible to study the luminosity-function of 
stars in globular clusters from the tip of the main-sequence down to the hydrogen burning limit and
observations are have been carried out for about a dozen galactic globular clusters (see for example
Paresce \& de Marchi 2000, Piotto \& Zoccali 1999 and references therein). Although 
some uncertainty in transforming the observed luminosity function into a mass-function remains, it seems
that most mass-functions can be fitted by power-laws with slopes comparable to or somewhat flatter 
then what one finds for the mass-function 
of low-mass stars in the galactic disc and bulge. There are indications that not all clusters
exhibit the same mass-function, although the reasons for the variation seen among different clusters are 
much
less clear. Differences might arise due to differences in the metallicities (De Marchi \& Paresce 1995)
or as a result of the dynamical evolution of the clusters (Piotto \& Zoccali 1999). 

However, in order to extract the scientifically much more interesting initial mass-function of stars 
from the observed mass-function, one has to know how
the IMF has changed due to the dynamical evolution of the clusters.
Furthermore, observations of stars in globular clusters usually cover only a limited range in radius, 
so the effect of mass-segregation of stars has to be understood in order to deduce the 
global mass-function from the observations. Although it is possible to correct for the effect
of mass-segregation by assuming equipartition and fitting multi-mass King-Michie models to the observed 
luminosity profile of a cluster (Gunn \& Griffin 1979), these question are probably best answered
by dynamical simulations.

The pioneering study in this field was made by Chernoff \& Weinberg (1990), who followed
the evolution of multi-mass clusters 
surrounded by a tidal cut-off and evolving under the combined influence of two-body relaxation and
mass-loss from stellar evolution by means of an Fokker-Planck code. They found that weakly 
concentrated clusters with a flat mass-function
are easily destroyed due to cluster expansion and tidal mass-loss resulting from the mass-loss by
stellar evolution. 
They also presented a number of observational parameters for clusters which survived up to the
present time. Their results were later extended by Fukushige \& Heggie (1995), Portegies Zwart et al.\ 
(1998), Giersz (2001) and Joshi et al.\ (2001). These studies generally confirmed the results
of Chernoff \& Weinberg for the cluster lifetimes, but showed that 
an isotropic Fokker-Planck code can give lifetimes significantly shorter than the results obtained by 
other methods for certain cases. Takahashi \& Portegies Zwart (1998, 2000) solved this discrepancy 
by showing that an anisotropic Fokker-Planck code with a better treatment of the external tidal field
gives good agreement with the $N$-body method. In their latter paper they conducted an extensive
survey of initial cluster conditions and showed that clusters can lose a large fraction of their
mass in a short time due to stellar evolution and still remain bound. Spurzem \& Aarseth (1996) also
found good agreement between various computational methods for the pre and post-collapse evolution of
a cluster containing 10000 equal-mass stars. 

Vesperini \& Heggie (1997) have performed a large set of $N$-body simulations of the evolution of
star clusters evolving under the combined influence of two-body relaxation, stellar evolution 
and disc-shocking. Their clusters 
started
with power-law mass-functions and they could establish a relation between the changes of 
the mass-function and the mass which is already lost from the clusters. They also showed that the slope of
the mass-function
near the half-mass radius will always stay close to the slope of the global mass-function. 
This result was later also obtained by Takahashi \& Lee (2000) from Fokker-Planck simulations.
Small-N simulations of multi-mass clusters in tidal fields were also carried out by de la Fuente 
Marcos (1995, 1996). His simulations showed that multi-mass clusters evolve considerably
faster than single-mass clusters and that the form of the initial mass-function and the fraction
of binary stars influences 
the lifetime of a star cluster.

Globular clusters evolve due to a number of internal and external processes, as for example
mass-loss due to stellar evolution, core-collapse and cluster re-expansion as a result of two-body relaxation 
and external tidal shocks from the parent galaxy. The realistic incorporation of all these effects into 
a dynamical simulation
has proven to be a substantial challenge for both hardware and software since the
different dissolution mechanisms act simultaneously and the range of
timescales involved in the numerical integration of the problem is rather large.
So far, realistic simulations of globular clusters were hampered by the fact that $N$-body
simulations with particle numbers realistic for globular clusters were not feasible, while
at the same time it was difficult to include all relevant physical effects into a Fokker-Planck or
Monte Carlo code with the necessary realism. Furthermore, as was shown by the 'Collaborative Experiment'
(Heggie et al.\ 1998) and Baumgardt (2001), the scaling of $N$-body simulations from small-$N$
clusters up to the globular cluster regime is not without risk, especially if many processes
acting on different timescales and depending on the number of cluster stars in differing ways,
are involved. 

With the completion of the GRAPE6 special purpose computers at Tokyo University
(Makino 2002), the situation has however changed dramatically. The GRAPE6 boards allow the
time-consuming calculation of the mutual gravitational forces to be done in hardware, thereby making it
possible to simulate the evolution of star clusters containing up to $3 - 5 \cdot 10^5$ stars with 
$N$-body programs like NBODY4 (Aarseth 1999) or KIRA (Portegies Zwart et al. 1999)
within a month of computing time. The number of stars feasible to simulate is thus comparable to what is
found in small globular clusters and the scaling of the results to real globular clusters,
if necessary at all, involves only very small factors, and is therefore less risky. 

In this paper we present results of a large set of $N$-body calculations for the dynamical evolution 
of globular clusters done on
GRAPE6. Our simulations include stellar mass-loss, two-body relaxation, 
and a realistic treatment of the external tidal field. We vary the particle number, the
initial concentration of the clusters and their orbits to see how these parameters affect the
dynamical evolution of the clusters. Our main focus is to study how quantities which can be checked
observationally, as e.g. the slope of the mass function or the fraction of cluster mass in compact 
remnants, change with time. From our results and observational data, we 
can draw conclusions about the starting conditions of individual clusters and the most likely form of the 
IMF of stars in globular clusters.

The paper is organised as follows: In chapter 2 we describe our simulation method and the initial 
set-up of our runs. Chapter 3 presents our results for the mass-loss of the clusters, the changes 
in their mass-functions and other cluster parameters. In this chapter we will also derive some simple
fitting formulas which describe the change of these parameters as a function of the fraction of time 
spent until complete dissolution.
In chapter 4 we will compare our results with 
observations of galactic globular clusters and in chapter 5 we finally draw our conclusions.

\section{Description of the runs}

We simulated the evolution of clusters containing between $N = 8.192$ and 131.028 (128K) stars,
increasing the particle numbers by successive factors of 2. The simulations were carried out
with the collisional Aarseth $N$-body code NBODY4 (Aarseth 1999) on the recently finished
GRAPE6 boards of Tokyo University. The NBODY4 code uses a Hermite scheme with individual
timesteps for the
integration and treats close encounters between stars by KS (Kustaanheimo \& 
Stiefel 1965) and chain regularisations (Mikkola \& Aarseth 1990, 1993). NBODY4 was specially
adapted to make use of the GRAPE6 hardware, which was used to calculate the gravitational forces
between the stars. 

Our clusters moved on circular or eccentric orbits through an external galaxy which followed a
logarithmic potential $\phi(R_G) = V_G^2 \ln R_G$ with 
circular velocity $V_G = 220$ km/sec. 
For the integration, we used a coordinate system where the cluster remained at the origin
and the center of the galaxy moved around the cluster, following the cluster orbit.
Such a coordinate system represents an accelerated but non-rotating one.
The force from the galactic center and the force due to the cluster acceleration were applied
to each star when it was advanced on its orbit through the cluster. 

In case of an eccentric orbit, we defined an eccentricity
parameter $\epsilon$ as $\epsilon = (R_{A}-R_{P})/(R_{A}+R_{P})$, where $R_{A}$
and $R_{P}$ are the apogalactic and perigalactic distances of the cluster. Most runs for 
eccentric orbits were made with $\epsilon = 0.5$. Runs were also made with
eccentricity parameters of $\epsilon = 0.2, 0.3, 0.7$ and 0.8 to study the influence
of orbital eccentricity on the lifetime of the cluster. 

Our clusters followed 
King profiles initially with central concentrations $W_0 = 5.0$ and 7.0. We also tried 
$W_0=3.0$ models, but found that they were susceptible to a quick disruption as a result of the stellar
mass-loss. This is in agreement with results of Fukushige \& Heggie (1995), who also found that
King $W_0 = 3.0$ models dissolve quickly unless the initial mass-function of the cluster stars
is very steep. We 
therefore decided to make no series of runs for $W_0=3.0$. Cluster radii were adjusted such that
the tidal radius of the King model was initially equal to the tidal radius of the external tidal
field, calculated according to
\begin{equation}
     r_t = \left( \frac{G \; m_c}{2 \; V_G^2} \right)^{1/3} \;\; R_G^{2/3}  \;\; ,
\label{rtide}
\end{equation}
where $m_c$ is the mass of the cluster, $V_G$ the circular velocity of the galaxy
and $R_G$ the distance of the cluster from the galactic center.
In case of an eccentric orbit, the cluster size was adjusted such that the tidal radius of the 
King model was equal to the tidal
radius at perigalacticon. 

Stars were removed from the simulation if their distance from the cluster center exceeded 
two tidal radii in case of a circular orbit, and two apogalactic tidal
radii in case of an eccentric orbit. All simulations were carried out until complete
cluster dissolution. During the runs, the cluster center was determined  
by using the method of Casertano \& Hut (1985). The number of bound stars and the tidal radius
of the cluster were then determined iteratively by first assuming that all stars still in the
calculation are bound and calculating the tidal radius with eq.\ 1. In a second step,
we calculated the mass of all stars inside $r_t$ and used it to obtain a
new estimate for $r_t$. This method was repeated until a stable solution was found. 
For clusters on eccentric orbits, the current distance from the center of the galaxy was used
to calculate the tidal radius in eq.\ 1.
Lagrangian radii were then determined from all stars
inside a sphere with radius $r_t$ around
the cluster center.

The initial mass-function of the
cluster stars was given by a Kroupa (2001) mass-function
with upper and lower mass-limits of 15 $\mbox{M}_\odot$ 
and 0.1 $\mbox{M}_\odot$ respectively. Within our mass-limits, this mass-function is a two stage
power-law with near Salpeter like slope above 0.5 $\mbox{M}_\odot$ and a shallower slope
below 0.5 $\mbox{M}_\odot$: 
\begin{equation}
     \xi(m) \; dm \sim
   \left\{ \begin{array}{l} 
       m^{-1.3} \; dm \;\; , \;\;\;\;\; m < 0.5 \; \mbox{M}_{\odot}\\ 
       m^{-2.3} \; dm \;\; , \;\;\;\;\; m \ge 0.5 \; \mbox{M}_{\odot}
    \end{array} \right. 
\end{equation}
Kroupa (2001) found this mass-function to be the mean mass-function of stars in the
galactic disc. The adopted slope at the low-mass end is also similar to what Zoccali et al.\ (2000)
found for the mass-function of low-mass stars in the galactic bulge. 
Our 
adopted mass-function leads to an initial mean mass of $<\!m\!> = 0.547 \; \mbox{M}_\odot$.

\input{table1.tex}

Stellar evolution was modeled by the fitting formulae of Hurley et al.\ (2000), which
give stellar lifetimes, luminosities and radii as a function of initial mass $m$ and metallicity $Z$ 
for all phases of stellar evolution from the zero age main-sequence to the remnant stages. To
apply the formulae, we assumed a metallicity of $Z=0.001$ for the cluster stars. Mass 
lost from stars due to stellar evolution was assumed to be immediately lost from the star cluster.
During the integration, the total mass lost due to stellar evolution was summed up since
it is an important reference quantity for our analysis. 

At the start of the simulation, our clusters did not contain binaries.
In our simulations, binaries and higher order hierarchies which
formed during the integration were treated as inert, assuming that no collisions or exchange of mass
take place and that the stellar evolution of the components
is not altered by their companions.

We did not apply kick velocities to stars which became neutron stars. Although 
there is 
evidence for the existence of such kick velocities (Hansen \& Phinney 1997), the exact
form of the kick velocity distribution is still a matter of debate. Applying kick 
velocities would mean that nearly all neutron stars are immediately lost from our clusters
since the sizes of the kick velocities exceed typical escape velocities from globular clusters by more than
an order of magnitude.
This seems to be equally unrealistic, given the fact that neutron stars are known to exist in 
globular clusters. As a matter of compromise 
we have therefore chosen a rather low value for the upper mass-limit of our adopted IMF, but 
keep all neutron stars which are formed during the run.

In order to perform the simulations, clusters were scaled from physical units to $N$-body units,
in which the constant
of gravitation, initial cluster mass and energy are given by $G=1$, $M=1$ and $E_C=-0.25$ respectively.
This scaling is straightforward since the total mass of the cluster together 
with the radius of the cluster orbit defines
the tidal radius and the crossing time in physical units. The stellar evolution timescales in
$N$-body units are obtained by requiring that their ratio to the crossing timescale must 
remain constant after rescaling the clusters to $N$-body units. When comparing our simulations
to observations, we assume an age of $T = 13$ Gyrs for the galactic globular cluster system.

Table 1 gives an overview of the simulations performed. It shows the number of cluster
stars $N$, the number $N_{Sim}$ of simulations for each cluster, the initial concentration and 
orbital type, the maximum distance from the galactic center and the initial mass and tidal radius
of the cluster. Also shown are the fraction
of mass lost due to stellar evolution, and the dissolution and core collapse times of the clusters. 
The dissolution times were defined to be the time when 95\% of the mass was lost from a cluster. 
Core collapse times were determined from the first minimum of the inner lagrangian radii.
For models for which more than one run was made, Table 1 gives the mean dissolution and core collapse 
times of all runs performed and the standard deviation of individual
runs around the mean. 

Family (I) represents our standard set of runs which are clusters moving on circular
orbits with $R_G=8500$ pc and following density profiles with concentration parameters $W_0 = 5.0$. 
The other cluster families are variations of this which
study the influence of the initial condition on the cluster evolution. Family (II) contains clusters
with higher central concentration $W_0 = 7.0$, and the following families contain clusters moving in eccentric orbits
with $R_{A} = R_G$. The next two families are clusters moving on circular orbits at smaller and larger 
galactocentric distances.
The last sets of simulations contain clusters on orbits with different eccentricities but otherwise
identical initial properties to family (I).

\section{Results}
 
\subsection{Dissolution}

\begin{figure*}
\epsfxsize=18cm
\begin{center}
\epsffile{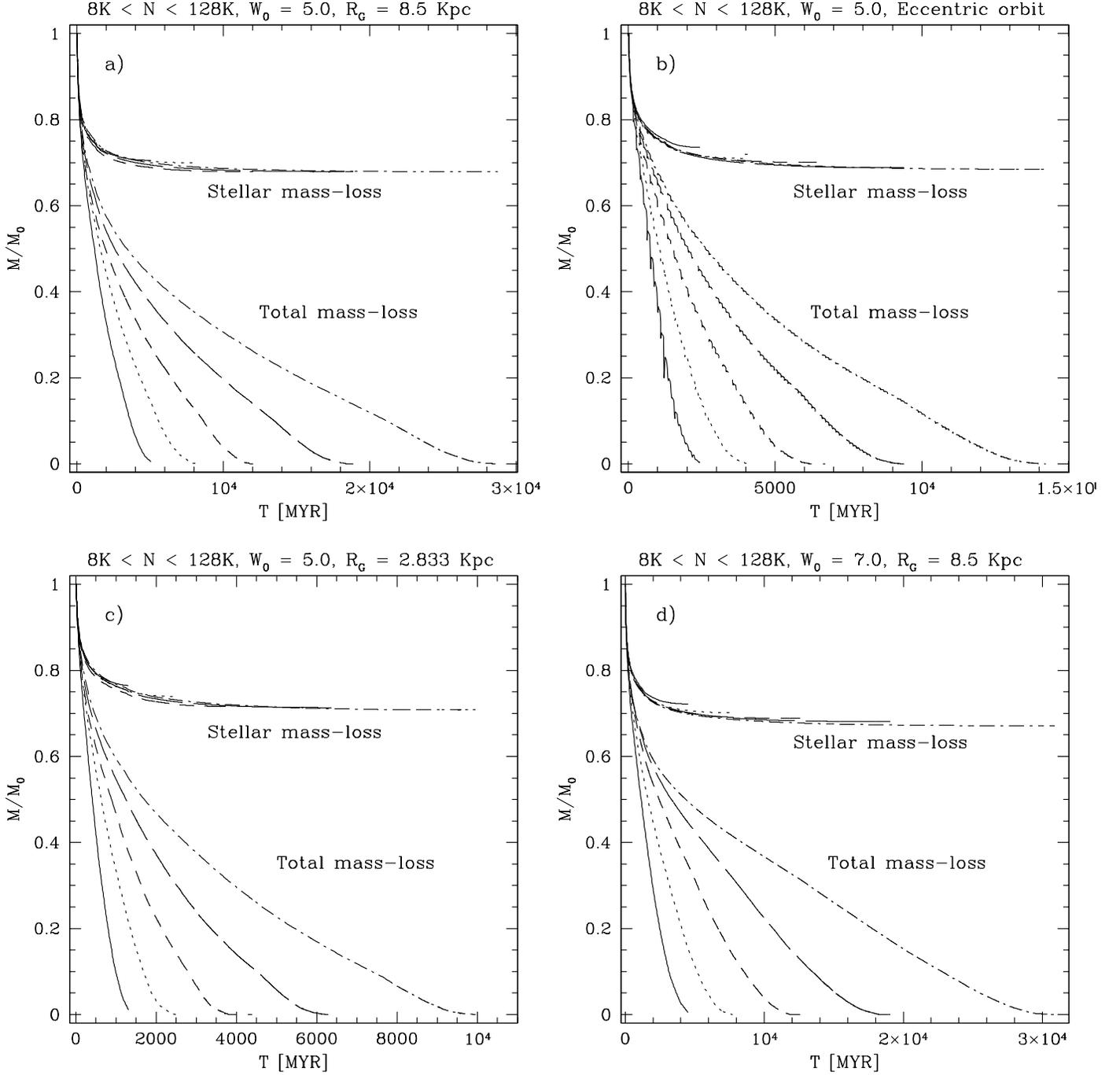}
\end{center}
\caption{Evolution of the bound mass with time for four different families of models: a) $W_0 = 5.0$, circular orbit
 with $R_G = 8.5$ kpc, b) $W_0 = 5.0$, eccentric orbit, c) $W_0 = 5.0$, circular orbit with $R_G = 2.833$ kpc,
  d) $W_0 = 7.0$, circular orbit. Clusters with 8K stars are shown by a solid line, 16K stars with
   a dotted line, 32K by a
   short dashed, 64K by a long dashed, and 128K by a dot dashed line. The upper curves in each panel
    show the stellar mass loss from bound stars, which reaches values around 30\% for most cases.}
\label{mlossa}
\end{figure*}

We start our discussion by presenting the results for the dissolution times.
Star clusters dissolve as a result of stellar evolution, two-body relaxation and
external tidal shocks, which all act on different timescales and scale differently with the particle
number. Stellar evolution is most efficient directly after cluster 
formation. Its influence on the dynamical evolution of star clusters diminishes thereafter since low-mass
stars retain a much larger fraction of their initial mass due to stellar evolution than high-mass
stars (see Weidemann 2000 and Fig.\ 18 of Hurley et al.\ 2000). Two-body 
relaxation arises from close encounters between cluster stars and leads to a slow diffusion of
stars over the tidal boundary. It acts on a timescale (Spitzer 1987)
\begin{equation}
     T_{rh} \sim \frac{\sqrt{m_c} \; r_h^{3/2}}{<\!m\!> \; G \; \ln (\gamma N)} \;\; ,
\label{trh}
\end{equation}
where $r_h$ is the half-mass radius of the cluster, $N$ the number of cluster stars and $\gamma$ the 
Coulomb logarithm. External tidal shocks happen in our simulations during pericenter passages of 
clusters on eccentric orbits.
Their strength does not change with the particle number and depends only on the density profile of the
cluster.

Fig.\ \ref{mlossa} depicts the evolution of bound mass with time for the first four families of cluster models. For
$N=8$K and 16K clusters, the average evolution is shown. The 
strong decrease of bound 
mass at the start is due to the mass-loss of high-mass stars from stellar evolution and levels off 
after about 25\% of
the cluster mass is lost. For clusters with $W_0 = 5.0$ there is some indication that the mass-loss
from stellar evolution is accompanied by some induced mass-loss resulting from an overflow of
stars over the tidal boundary. This mass-loss is much reduced in case of the more concentrated $W_0=7.0$
models. It can be seen in Fig.\ 1 and Table~1 that the total mass lost due to stellar evolution 
of the bound stars is almost the same in all cases, since we usually obtain values around 30\%.  Most 
discrepant are small-$N$ clusters 
moving at galactocentric distances of $R=2.8$ kpc, whose lifetimes are comparable with the 
timescale of stellar evolution.
If we take a population of stars which initially follow our adopted IMF and 
evolve them with the Hurley et al.\ (2000) stellar evolution routines, we find that 33.1\% of the mass is 
lost due to stellar evolution after $2.0 \cdot 10^4$ Myr. This is quite close to the values found in 
most of our runs. Mass loss due to stellar evolution happens therefore very early in the evolution and is
nearly finished by the time the other mass-loss mechanisms become important for the mass-loss.

\begin{figure}
\epsfxsize=8.3cm
\begin{center}
\epsffile{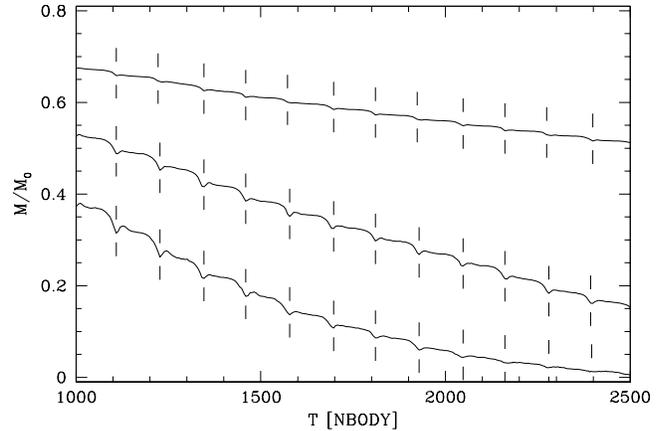}
\end{center}
\caption{Evolution of bound mass for clusters on eccentric orbits. Shown is the evolution for clusters with (from
   bottom to top) $N=8$K, 16K, 128K stars on $\epsilon=0.5$ orbits. Solid lines mark pericenter passages. During these
   passages, the number of bound stars drops sharply as a result of the shrinking tidal radius. In the intermediate 
   phase between two pericenter passages, the number of cluster stars changes only slowly. The fractional mass-loss
   per passage is strongest for low-$N$ clusters.}
\label{mloss_ecc}
\end{figure}

While clusters on circular orbits lose mass fairly smoothly, clusters on eccentric orbits
lose mass in distinct steps as stars escape mainly while the cluster is near perigalacticon
(see Fig.\ \ref{mlossa}b). This is shown in greater detail for 3 clusters on $\epsilon=0.5$ orbits 
in Fig.\ \ref{mloss_ecc}. During each perigalactic passage, stars at or beyond the perigalactic tidal 
radius receive a strong velocity kick from the varying tidal field and most are stripped away from the 
clusters, causing a decrease in bound cluster mass.
As the clusters move outward again, their tidal radii expand and the clusters can fill a 
much larger volume of space. Clusters will therefore undergo an expansion which is
driven by relaxation, although only few stars move over the tidal boundary during this phase.
The next perigalactic passages removes most stars scattered 
beyond the perigalactic tidal radius. One would expect that the fraction of stars 
removed at each passage should decrease with increasing $T_{RH}/T_{ORB}$, or, if the orbital timescale is constant,
with the particle number. Fig.\ \ref{mloss_ecc} shows that this is indeed the case. 

\begin{figure}
\epsfxsize=8.3cm
\begin{center}
\epsffile{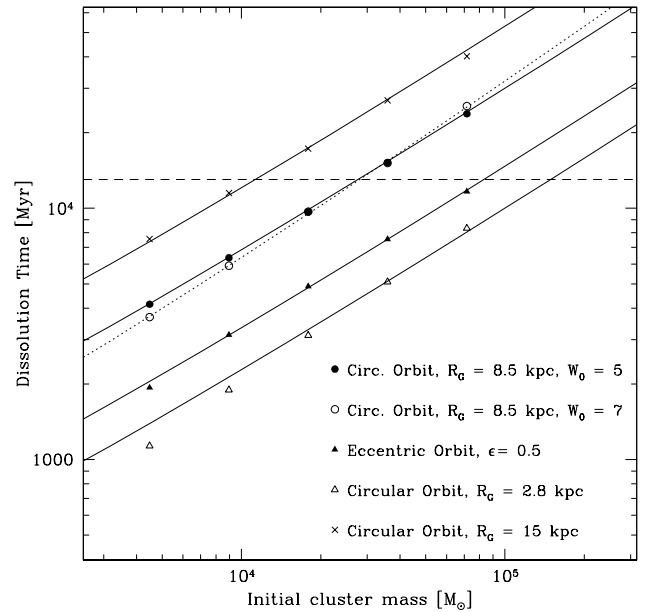}
\end{center}
\caption{Dissolution times as a function of the initial cluster mass and orbital type.
 The horizontal dashed line shows an age of 13 Gyrs. Solid lines show a scaling
 proportional to $T_{RH}^{0.75}$, which provides a good fit for King $W_0=5.0$ clusters. 
 The dotted line is a scaling law proportional to $T_{RH}^{0.82}$, which fits the 
 lifetimes of the King $W_0=7.0$ clusters.} 
\label{tdiss}
\end{figure}

Fig.\ \ref{tdiss} shows the cluster lifetimes as a function of the initial mass and orbital
type. For the same orbit, King $W_0=5.0$ clusters have lifetimes very similar to the King
$W_0=7.0$ ones, the difference never exceeding 10\%. The initial concentration
seems to influences the lifetimes only very little, provided clusters survive
the initial phase when mass-loss from stellar evolution dominates.

The horizontal line in Fig.\ \ref{tdiss} shows the age of the cluster system, which we
assume to be 13 Gyrs. At the solar radius, clusters with mass less than $M=2.8 \cdot 10^4 \; \mbox{M}_\odot$ are
completely destroyed within one Hubble time. This value rises to $8.1 \cdot 10^4 \; \mbox{M}_\odot$ in case
of an 
eccentric orbit with $\epsilon=0.5$, and $1.49 \cdot 10^5 \; \mbox{M}_\odot$ for a cluster at $R_G = 2.833$ kpc
from the galactic center. Several clusters in Table~1 have lifetimes well exceeding a Hubble time.
The cluster starting with $N=128$K stars, $W_0=5.0$, and moving on a circular
orbit with $R_G = 8.5$ kpc for example still has a bound mass of $M = 1.7 \cdot 10^4 \mbox{M}_\odot$ after $T=13$ Gyr and
would have evolved into a small globular cluster.

With the exception of clusters moving at $R_G = 2.8$ kpc, we always obtain scaling laws $T \sim T_{rh}^x$ 
with $x$ equal to 0.75 for King $W_0=5.0$ clusters moving on similar orbits. King $W_0$ = 7.0
clusters show a slightly steeper scaling law with $x=0.82$. Both values are comparable to 
what Baumgardt (2001, 2002) found for single-mass clusters in tidal
fields, and also close to the scaling laws obtained by the 'Collaborative Experiment' (Heggie et al.\ 1998) for
multi-mass clusters without stellar evolution. The dependence of the lifetimes on the relaxation time
is weaker than linear since potential escapers need a considerable amount of time to
escape from clusters in tidal fields (Fukushige \& Heggie 2000). While they remain trapped to the
cluster, they experience further scatterings
with other cluster stars, which complicates the escape process and leads to a deviation from
a linear scaling with the relaxation time.
\begin{figure}
\epsfxsize=8.3cm
\begin{center}
\epsffile{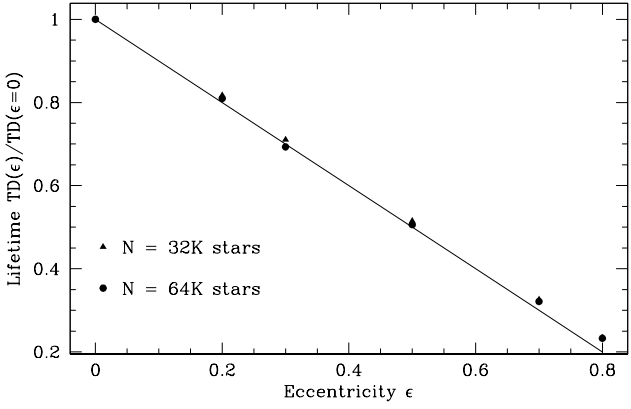}
\end{center}
\caption{Lifetimes of clusters moving on orbits with different eccentricities $\epsilon$ but the same apogalactic
 distances. Lifetimes are divided by the lifetime of
  a cluster moving on a circular orbit with radius equal to the apogalactic radius of the clusters
  on the eccentric orbits. The solid line shows a relation $(1-\epsilon)$, which provides
   a satisfactory fit for all eccentricities.}
\label{elldepf}
\end{figure}

Following Baumgardt (2001, eq.\ 16), we expect the dissolution time of a cluster to be given by:
\begin{equation}
T_{DISS} = k \; T_{RH}^x \; T_{CROSS}^{1-x} \;\;. 
\end{equation}
Using eq.\ \ref{trh} and the fact that the crossing time is proportional to $r_h^{3/2}/\sqrt{G\;M}$, 
we can rewrite this as
\begin{equation}
T_{DISS} = k' \left( \frac{N^{1/2} r_h^{3/2}}{G^{1/2} m^{1/2} \ln(\gamma N)} \right)^x  
 \! \left( \frac{r_h^{3/2}}{G^{1/2} N^{1/2} m^{1/2}}\right)^{1-x} \!\!\!\! ,
\label{step}
\end{equation}
where $m$ is the mean mass of a star in the cluster. 
If we assume that all radii in a cluster scale with its tidal radius, $r_h$ is proportional
to $r_t$, we can rewrite eq.\ \ref{step} with the help of eq.\ \ref{rtide} to obtain
\begin{equation}
T_{DISS} = k' \left( \frac{N}{\ln(\gamma N)} \right)^x \frac{R_G}{V_G} \;\;\; , 
\end{equation}
or, equivalently 
\begin{equation}
\frac{T_{Diss}}{\mbox{[Myr]}} \; = \; \beta \; \left( \frac{N}{ln(\gamma \, N)} \right)^x \; 
  \frac{R_G}{\mbox{[kpc]}} \; \left( \frac{V_G}{220\; \mbox{km/sec}} \right)^{-1}\,\, . 
\label{eltime}
\end{equation}
The
exact value of the factor $\gamma$ in the Coulomb logarithm depends on the mass-spectrum
and is rather uncertain for multi-mass clusters. In addition, it is almost certainly
varying in the simulated clusters since the mass-spectrum itself changes during the evolution.
Fortunately, due to the large number of cluster stars, our results are rather
insensitive to the exact value of $\gamma$, so we can adopt the value found by Giersz \& Heggie (1996)
for their multi-mass clusters: $\gamma=0.02$. From a fit to the lifetimes of the $W_0 = 5.0$ clusters in circular orbits, we
then obtain $x = 0.75$, $\beta = 1.91$. The corresponding values of $x$ and $\beta$
for $W_0 = 7.0$ clusters are $x=0.82$ and $\beta = 1.03$.
\begin{figure}
\epsfxsize=8.3cm
\begin{center}
\epsffile{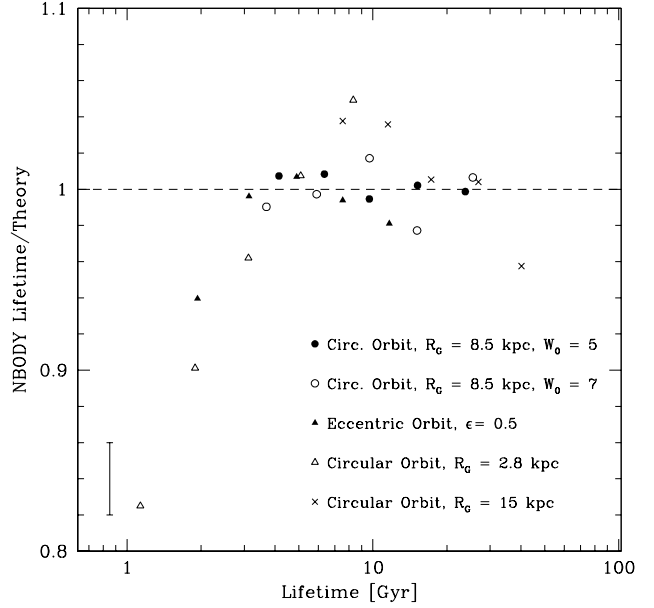}
\end{center}
\caption{Ratio of $N$-body to predicted lifetimes according to eq.~\ref{gtime} 
   as a function of the $N$-body lifetimes for the same clusters as in Fig.\ \ref{tdiss}. The errorbar gives
  the uncertainty of a single calculation, assumed to be 2\%. For most clusters, no systematic deviation exists
  between our prediction and the $N$-body results. The exception are clusters dissolving in less than
  3 Gyrs, whose lifetimes are comparable to the timescale of stellar evolution.} 
\label{tdiff_pap}
\end{figure}

Despite the fact that clusters on eccentric orbits are subject to external tidal shocks,
their lifetimes can also be fitted by a scaling proportional to $T_{RH}^{0.75}$, similar to the circular case
(see Fig.\ \ref{tdiss}). Hence, in order to predict lifetimes
we only have to determine how the dissolution time of a cluster depends on 
the eccentricity of its orbit. In order to do so, we performed a series of runs of clusters with identical
initial parameters, but moving on orbits with different eccentricities $\epsilon$, 
starting at similar apogalactic distances (families VI to IX of Table~1).
Fig.\ \ref{elldepf} depicts how the dissolution times of these clusters change as a function of orbital
eccentricity. The solid line shows a relation according to 
\begin{equation}
T_{DISS}(\epsilon) = T_{DISS}(0) \cdot (1-\epsilon)  \;\;\;\; ,
\label{elldepe}
\end{equation}
which provides a good fit to the $N$-body results for both particle numbers. Here $T_{DISS}(0)$ is the lifetime of a
cluster
moving on a circular orbit with radius similar to the apogalactic radii of the eccentric orbits. Eq.\ \ref{elldepe}
will be invalid when the predicted lifetime of a cluster becomes comparable to its orbital timescale.
The corresponding relation for similar perigalactic distances would be
\begin{equation}
T_{DISS}(\epsilon) = T_{DISS}(0) \cdot (1+\epsilon)  \;\;\;\; ,
\label{elldep2}
\end{equation}
i.e. the lifetime of a cluster on an eccentric orbit can be increased by at most a factor of 2 compared
to the lifetime of a cluster moving on a circular orbit at the same perigalactic distance. A similar, relatively
small increase was also found by Joshi et al.\ (2001) in Monte Carlo simulations of a cluster on an $\epsilon=0.6$
orbit.

Summarizing, the lifetimes of clusters starting with a Kroupa (2001) IMF and moving
on orbits with eccentricity $\epsilon$ through a logarithmic external potential are given 
by the following formula: 
\begin{equation}
\nonumber \frac{T_{Diss}}{\mbox{[Myr]}} = \beta \left( \frac{N}{ln(0.02 \, N)} \right)^x
  \!\! \frac{R_G}{\mbox{[kpc]}} 
  \left( \frac{V_G}{220 \mbox{km/sec}} \right)^{-1} \!\!\!\! (1-\epsilon)\, ,
\label{gtime}
\end{equation}
with the values of $\beta$ and $x$ depending on the initial concentration.
Fig.\ \ref{tdiff_pap} shows the difference between the $N$-body lifetimes and our prediction for the clusters
of Fig.\ \ref{tdiss}. The typical uncertainty of a single calculation, based on the values in 
Table 1, seems to be roughly 2\%. Most $N$-body results are therefore compatible with our prediction. 
Most discrepant are small-$N$ clusters on $R_G = 2.8$ kpc orbits, which dissolve considerably faster than predicted.
The reason is that they contain high-mass stars for a much larger fraction of their lifetime.
Since high-mass stars are very effective in scattering low-mass stars out of a cluster, the cluster
lifetimes become smaller. Nevertheless, our prediction is accurate to within a few percent for most cases. 
Chapter 4.1 we will compare our formula with some results from the literature and discuss its implication for the 
depletion of the galactic globular cluster system. 
\begin{figure}
\epsfxsize=8.3cm
\begin{center}
\epsffile{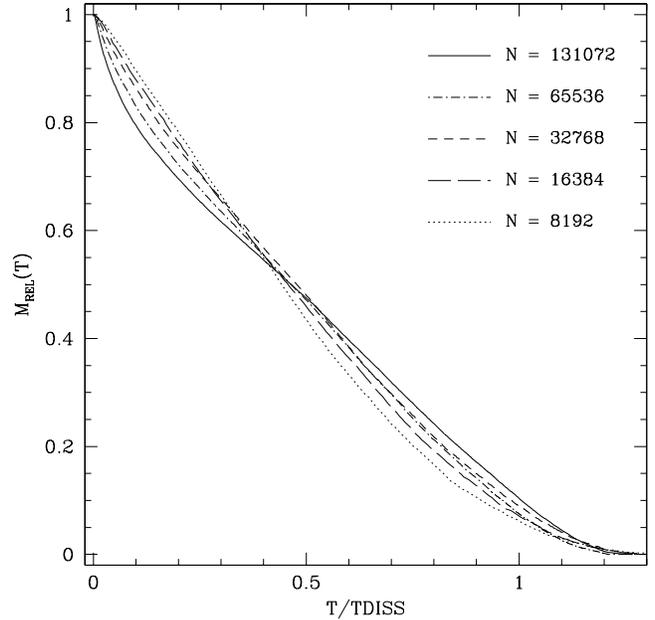}
\end{center}
\caption{Evolution of mass with time for the $W_0=7.0$ clusters after the mass lost due to stellar
 evolution has been subtracted from the total mass loss and the timescales have been divided by the
 dissolution times from eq.\ \ref{eltime}. Clusters lose mass roughly linearly until shortly before the end.}
\label{ml_scaled}
\end{figure}

We finally discuss the mass-loss rate. Fig.\ \ref{ml_scaled} shows how the cluster mass changes with time 
after the mass-loss curves were normalized 
by the mass not lost due to stellar evolution, i.e. 
\begin{equation}
 M_{REL}(T)  = \frac{M(T)}{M(0) - M_{SEV}(T)} \,\, , 
\end{equation}
where $M(T)$ and $M_{SEV}(T)$ are the cluster mass and cumulative mass lost due to stellar evolution
at time $T$ respectively. Shown is the case $W_0=7.0$
which has the smoothest mass-loss curve. We have normalized
the physical time by the lifetime of each cluster calculated according to eq.\ \ref{eltime}. The differences 
among the different curves are relatively small and they are all fairly close to a straight line 
approximation for most part of the evolution. Only near the end, a leveling off of the mass-loss 
can be seen. A linear dependence should be a good approximation to the mass-loss curve, since, 
as a cluster loses mass, its tidal radius shrinks proportional to $r_t \sim M^{1/3}$, and,
assuming a self-similar behavior, all other radii shrink in a similar way. Since the relaxation 
time depends on $r$ and $M$ roughly as $T_{RH} \sim \sqrt M \; r^{3/2}$, $T_{RH}$
is proportional to the mass left in the cluster. From the results obtained so far, we would
expect the mass-loss rate to be given by
$d M/d T \sim -M/T_{RH}^x = -M^{1-x}$, which varies only slowly with $M$ since $x$ is not too
far from unity.

Hence, if we assume that the mass-loss due to stellar evolution takes place immediately after cluster
formation, and that later on globular clusters lose mass linearly, the mass still bound to a 
globular cluster with initial mass $M_0$ can be approximated as 
\begin{equation}
M(T) = 0.70 \, M_0 \, (1-T/T_{Diss}) \;\;\;\;\; ,
\label{emfunc}
\end{equation}
for $T < T_{Diss}$.

\subsection{Evolution of the mass-function}

\begin{figure*}
\epsfxsize=18cm
\begin{center}
\epsffile{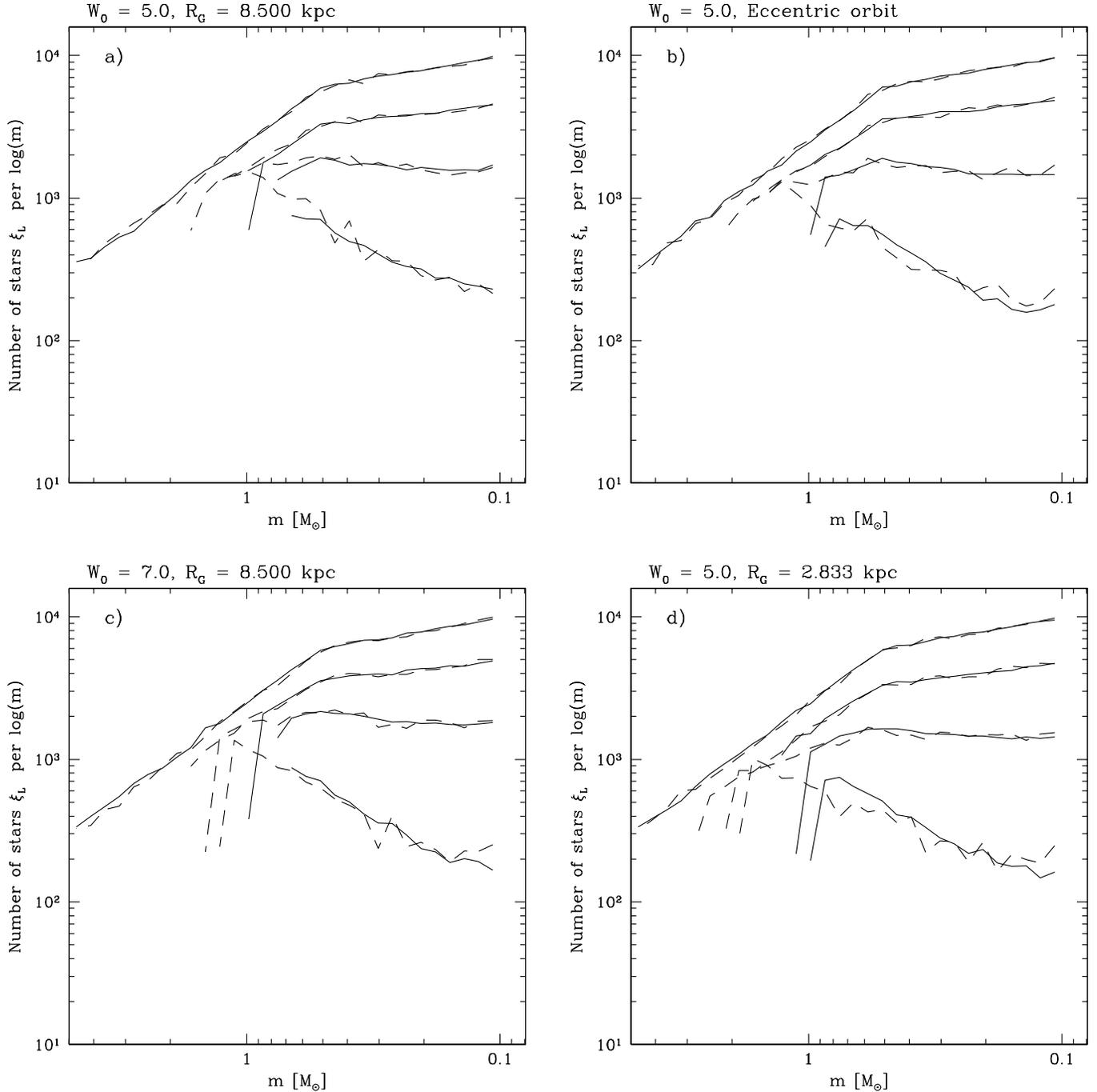}
\end{center}
\caption{Combined mass function of main-sequence stars and giants for different orbital families. Shown are the
 mass-functions at the start, and when 30\%, 60\% and 90\% of the cluster lifetime are over. Solid
 lines are for the biggest simulated clusters of each family ($N=128$K), dashed lines show the sum of the
 four smallest clusters
 with $N=8$K stars. Small-$N$ clusters were shifted to contain the same number of low mass stars
  ($m<0.5 \; \mbox{M}_\odot$) than the high-$N$ runs. The preferential depletion of 
 low-mass stars leads to mass-functions which decrease towards small $m$ towards the end.
  The details of this depletion are nearly independent of the initial particle number and orbital type.}
\label{allmf}
\end{figure*}

Fig.\ \ref{allmf} depicts the evolution of the combined mass-function of main-sequence stars and giants as a 
function of time. The bend in the Kroupa (2001) IMF at $m=0.5 \; \mbox{M}_\odot$ can clearly be seen. It survives 
until fairly late but is difficult to distinguish by the time 90\% of the mass is lost.
Below $m=0.5 \; \mbox{M}_\odot$, the mass function can be characterised by a single power-law
throughout most part of the evolution. Only for clusters on eccentric orbits, some deviations from a
power-law occur near the end of the lifetime at the low-mass end.
There are indications that the mass-function of some galactic globular
clusters cannot be characterised by single power-law mass-functions (de Marchi, Paresce \& Pulone 2000). If true, 
these features are likely to be primordial, since Fig.\ \ref{allmf} shows that dynamical evolution does not
lead to the formation of such features.
 
Fig.\ \ref{allmf} also shows the preferential depletion of low-mass stars as a result of mass-segregation
and overflow over the tidal boundary. The depletion already sets in before 30\% of the lifetime 
is over. After about 60\% of the lifetime has passed, low-mass stars are depleted to such an extent
that the number of cluster stars starts to drop towards the low-mass end. 

Differences between large and small-$N$ 
models are relatively small for any orbital family if we compare the mass-functions at the same fractional lifetime,
despite the fact that there is already a factor of 16 difference in the particle numbers. It should therefore be safe
to apply our results also to globular clusters, which mostly have particle numbers within 
a factor of 10 of our largest runs.
Furthermore, differences between clusters of different orbital families are also small,
indicating that the mass-function changes in an universal way, independent of particle number and orbital type
of the parent cluster. It is often stated in the literature that the preferential depletion of low-mass stars 
is due to mass-segregation and external tidal shocks. However, our simulations show that relaxation is 
responsible for the depletion of low-mass stars, since we get the same depletion whether clusters are subject to 
external tidal shocks (as in Fig.\ \ref{allmf}b) or not. 

\begin{figure}
\epsfxsize=8.3cm
\begin{center}
\epsffile{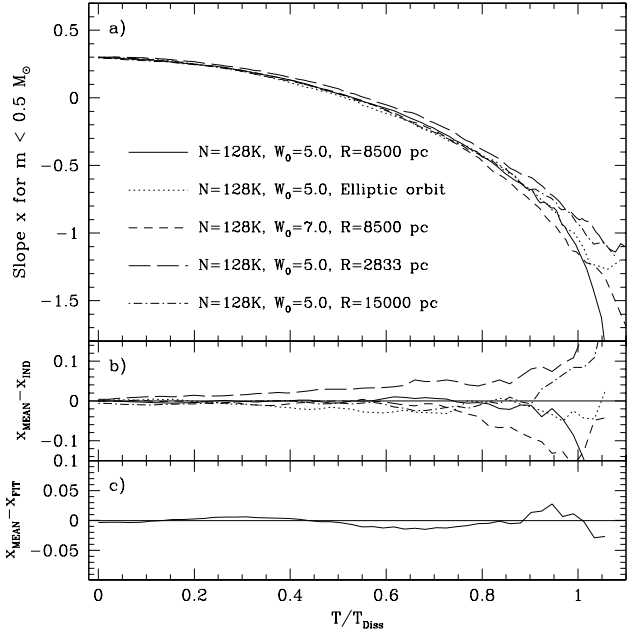}
\end{center}
\caption{Slope of the mass function of low-mass stars as a function of the fraction of time spent until 
  complete cluster dissolution. a) Slopes for individual $N$-body runs. b) Differences between the 
  individual slopes and an average taken from all runs shown in panel a). The differences are small
  and could entirely be due to finite-$N$ scatter. c) Difference of the mean from a least-square
  fitting curve according to eq.\ \ref{mftf}.}
\label{mfslt}
\end{figure}

Fig.\ \ref{mfslt} depicts the evolution of the slope of the mass function at the low-mass end in dependence 
of the time elapsed until cluster dissolution. The slope was determined from 
a fit to the distribution of stars with masses $m \le 0.5 \; \mbox{M}_\odot$. Note that we have put stars 
into bins equally 
spaced in $\log m$, so the slope $x$ of the mass-function is by one smaller than in case of a plot
over $m$, i.e. $x=\alpha-1$. Fig.\ \ref{mfslt} confirms that the slope of the mass-function changes in all
clusters in more or less the same way. During the first 20\% of the cluster lifetime, the mass-function
remains nearly constant, which might arise from the fact that in the beginning masses of
stars with $m \le 0.5 \; \mbox{M}_\odot$ are fairly small
compared to the mass of the most heavy stars in the cluster, hence they will be depleted at similar rates. 
Later, the slope changes with increasing speed since low-mass stars become more and more depleted.
Differences between individual runs (Fig.\ \ref{mfslt}b) remain 
small and could entirely be due to statistical errors, which are of order $\Delta x = 0.02$
at the start of the runs and increase as the number of cluster stars decreases. 

We can fit the mean evolution by a formula like:
\begin{equation}
 x = 0.3-1.51 \left(\frac{T}{T_{Diss}}\right)^2 + 1.69 \left(\frac{T}{T_{Diss}}\right)^3 - 
  1.50 \left(\frac{T}{T_{Diss}}\right)^4 \;,
\label{mftf}
\end{equation}
where we have chosen the functional form first and determined the coefficients from a least mean-square
fit to the $N$-body results. 
The differences between eq.\ \ref{mftf} and the mean curve from the $N$-body results are small and are
shown in Fig.\ \ref{mfslt}c. Eq.~\ref{mftf} should therefore give an accurate prediction for the 
change of the stellar mass-function in a globular cluster, provided its dissolution time is known. 
\begin{figure}
\epsfxsize=8.3cm
\begin{center}
\epsffile{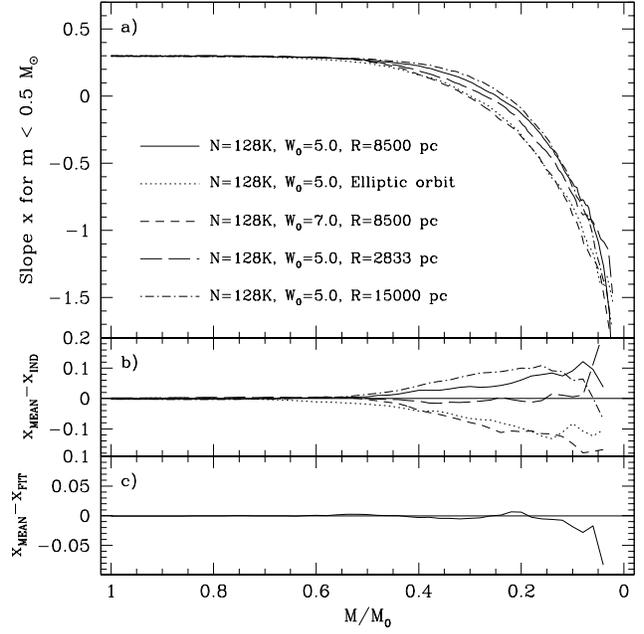}
\end{center}
\caption{Same as Fig.\ \ref{mfslt} but now in dependence of the mass fraction which is still bound
  to the clusters.}
\label{mfsl}
\end{figure}
\begin{figure*}
\epsfxsize=17cm
\begin{center}
\epsffile{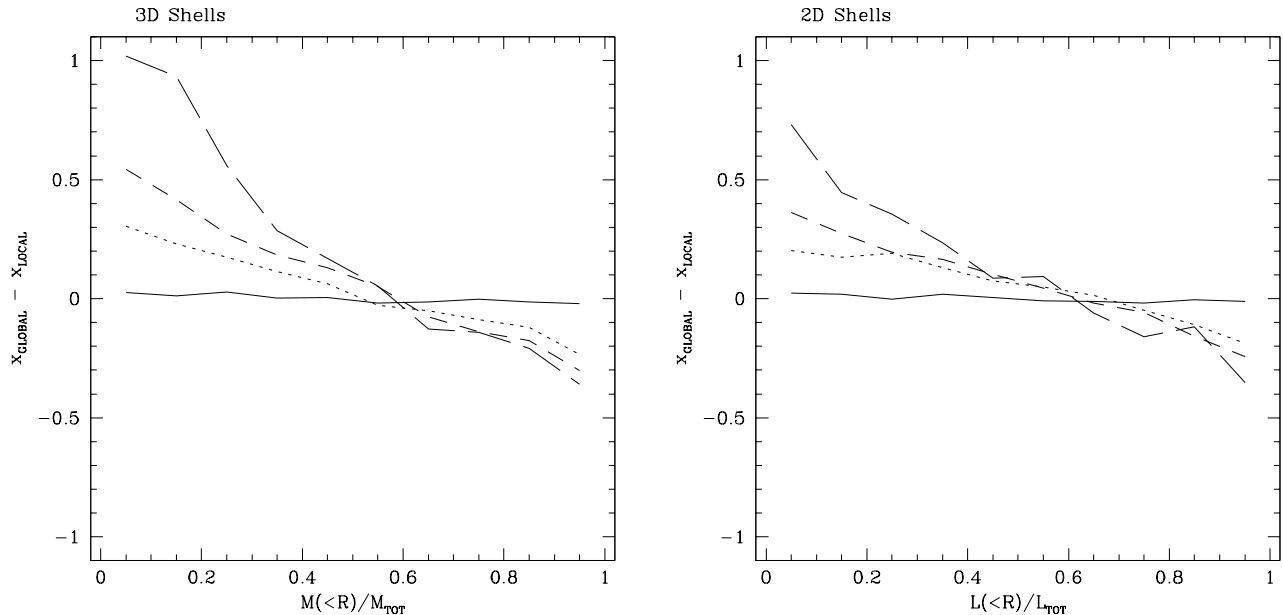}
\end{center}
\caption{Difference of the local mass-function minus the global one at four different times for the
 run with $N=128$K stars and $R_G=8.5$ kpc. Shown are differences at the start of the calculation
 (solid line) and when 30\% (dotted), 60\% (short dashed) and 90\% (long dashed) of the
  lifetime have passed. The left panel shows differences for
  3D mass-shells, the right panel shows differences in projection and as a function of the enclosed
 fraction of cluster light.
  Inside the half-mass radius, local mass-functions are steeper than the
  global one as a result of mass-segregation. Throughout the run, mass-segregation is increasing in the
  inner parts. Differences in projection are smaller than the 3D ones.
  Best agreement with the global mass-function is achieved just outside the half-light radius.}
\label{mfslo}
\end{figure*}

Similar results can also be found for the dependence of the slope of the mass-function 
on the fraction of mass that is already lost from a cluster (Fig.\ \ref{mfsl}). In this case, the slope does not change
significantly until nearly half the cluster mass is lost since in the beginning most mass is lost due to stellar
evolution. We can fit the mean evolution by:
\begin{equation}
 x = 0.3-\exp(0.67-6.19 M/M_0-3.24 \left(M/M_0\right)^2)
\end{equation}
The lower panels of Fig.\ \ref{mfsl} show the difference between the mean curve and individual runs and the difference
between a mean curve and our fitting function. Differences are of the same size as before.

Mass-segregation changes the local mass-function relative to the global one since stars heavier than
the mean sink towards the center and lighter ones move outward. If uncorrected for, this could bias 
the observed mass-function, as observations usually cover only a limited range in radii, most often 
around the half-mass radius. In order to determine to which extent mass-segregation changes local mass-functions,
we calculated local mass-functions from all stars with $m<0.5 M_\odot$ at different radial shells and compared them 
with the
global one. Fig.\ \ref{mfslo} shows the results for the $N=128$K star cluster from our standard family.
Shown are differences for 3D shells (panel a) and (projected) 2D shells (panel b).
In order to allow the most easy comparison with observations, projected shells consider the emitted fraction
of cluster light. 

In the beginning, the mass-functions are the same at any radius since our clusters start without 
mass-segregation. Mass-segregation is however quickly established and one finds more negative slopes for 
the mass-function in the inner parts since heavier stars sink into the center and lighter stars
move outwards.
The largest differences are found in the center and amount to about $\Delta x = 1.0$ in the 3D case
close to the end. The reason for the strong difference compared to the average slope over all radii is 
that low-mass stars are almost completely absent in the cluster center. 
Outside the half-mass radius, local mass-functions differ 
much less from the global one and are always shallower. As one would expect, differences are smaller if viewed in
projection (panel b). The radii containing between 50-80\% of the emitted cluster light
show the smallest difference from the global mass-function. For the cluster shown, this amounts typically to
distances of 1 to 1.8 (projected) half-light radii. Similar results are obtained for the other clusters. 
Mass-segregation is stronger and sets in faster in clusters undergoing core-collapse earlier in their 
evolution, as e.g. clusters starting from $W_0 = 7.0$ models. Here differences between local and 
global mass-functions are about $\Delta x = 0.2$ larger in the center than in the $W_0 = 5.0$ case.

\subsection{Cluster composition and mass-to-light ratio}

The question to which extent globular clusters contain dark matter is still a matter of debate 
(see Heggie \& Hut 1996). Dark matter could be present either in the cluster centers in the form
of intermediate mass black holes (Gerssen et al. 2002, Gebhardt, Rich \& Ho 2002), or in the form of 
faint stellar mass white dwarfs
and neutron stars. Previous studies (Vesperini \& Heggie 1997, Giersz 2001) have shown that the fraction
of white dwarfs is rising in a star cluster as it evolves. According to Giersz (2001), more than half the 
total cluster mass can be 
in the form of white dwarfs close to the end of its lifetime, depending on the initial mass-function and 
cluster lifetime. Since our initial mass-function and the results for the cluster lifetimes are different from 
previous studies, it might be interesting to look how the mass fraction of compact remnants changes 
in our clusters. 
 
\begin{figure}
\epsfxsize=8.3cm
\begin{center}
\epsffile{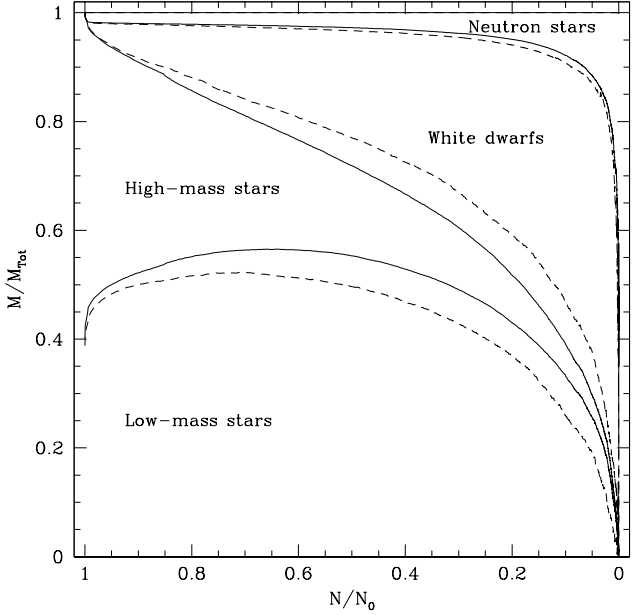}
\end{center}
\caption{Fraction of mass in different stellar groups in dependence of the number of 
   stars still bound to the cluster. Solid lines show the evolution for the run with $N=128$K stars, 
  $W_0 = 5.0$ and $R_G = 8.5$ kpc. 
   The dashed lines show the evolution for a cluster with $N=32$K stars but otherwise similar initial condition.
   The fraction of compact remnants increases throughout the evolution. The mass-fraction of low-mass stars increases
   in the beginning due to the mass-loss of high-mass stars. Later their fraction decreases as a result of the
   preferential depletion of low-mass stars.} 
\label{mdist}
\end{figure}

Fig.\ \ref{mdist} depicts for two different runs the fraction of mass in different stellar groups as
a function of the number of stars left in the cluster. 
We have divided main-sequence stars and giants into two categories,
high and low-mass stars, depending on whether their mass exceeds 0.7 $\mbox{M}_\odot$ or not. In the beginning, 
most mass in the cluster can be found in high-mass main-sequence stars. Their mass-fraction 
drops steadily since stellar evolution constantly transforms the high-mass end of the 
mass-function into compact remnants. 

The mass-fraction of low-mass stars increases initially due to the mass-loss of the high mass stars. 
Later, their mass-fraction decreases since they
are the lightest mass component and are therefore preferentially removed from the cluster.

The decrease in the mass-fraction of main-sequence stars is compensated for 
by a constant increase of the mass-fraction of white dwarfs and, very close to the end,
also by an increase in the mass-fraction of neutron stars. 
For the cluster with $N=128$K stars initially, white dwarfs become the dominant mass group
after roughly 80\% of the cluster stars are lost. The mass and number
fractions of giants are always very small and they are therefore
not shown separately. The dotted lines show the evolution for a cluster with a smaller initial particle 
number. It can be seen that the overall evolution is similar, except that the fraction of white dwarfs is
smaller since in a cluster with shorter dissolution time a smaller fraction of stars is turned
into white dwarfs due to stellar evolution. 

\begin{figure}
\epsfxsize=8.3cm
\begin{center}
\epsffile{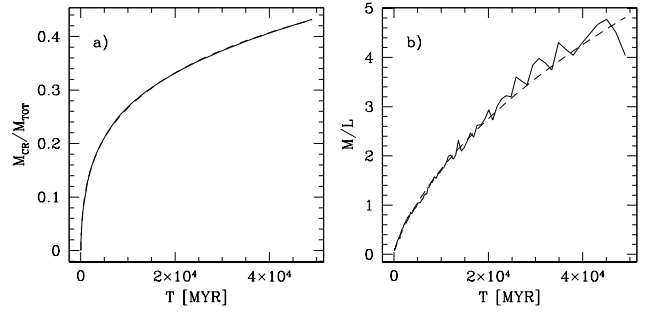}
\end{center}
\caption{Evolution of the mass-fraction of compact remnants (left) and the cluster mass-to-light ratio (right) 
 for a
 population of 500000 stars following a Kroupa (2001) IMF and evolving under the influence
  of stellar evolution alone. Solid lines represent results of simulations, dashed lines the fit according to
  eqs.\ \ref{fcrsev} and \ref{mlrel}.}
\label{sev}
\end{figure}

For a quantitative understanding of how the fraction of mass belonging to the different mass-groups 
changes, one first has to know the rate with which stellar evolution turns mass into compact remnants.
Fig.\ \ref{sev} a) depicts how the combined mass-fraction of white dwarfs and neutron stars increases with time
if we take a population of 500000 stars that follow a Kroupa IMF initially and evolve them
with the Hurley et al.\ (2000) stellar evolution routines.
For $T>10$ Myr, we can fit the whole evolution rather accurately by the following formula: 
\begin{eqnarray}
\nonumber \left. F_{CR} \right|_{SEV} & = 0.024 - 0.048 \; z + 0.0241 \; z^2 - 0.00055 \; z^3 \\ 
 & + 0.00033 \; z^4 \; ,
\label{fcrsev}
\end{eqnarray}
where $z = \log_{10}(T)$. Here $T$ is measured in Myrs and the coefficients were determined by a least 
square fit to the evolution of the test population. The differences 
between our fitting formula and the real evolution always remain 
smaller than $\Delta F$ = 0.005.
\begin{figure}
\epsfxsize=8.3cm
\begin{center}
\epsffile{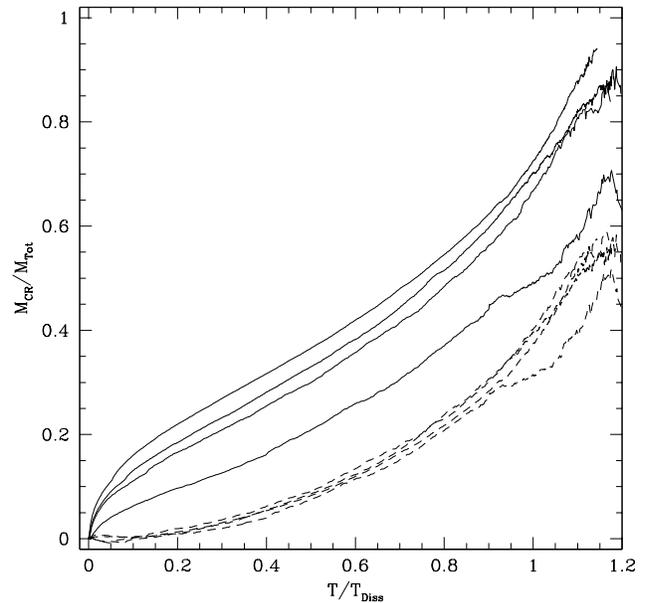}
\end{center}
\caption{Fraction of mass in compact remnants for 4 individual runs. From top to bottom: $N=128$K, $W_0=5.0, R_G=8.5$
  kpc;
  $N=64$K, $W_0=7.0$; $N=128$K, eccentric orbit; $N=32$K, $W_0=5.0, R_G=2.8$ kpc. Solid lines
   show the evolution before, dashed lines after the mass-fraction due to stellar evolution (eq.\ \ref{fcrsev}) is
   subtracted from the runs. After this subtraction, the evolution is rather similar in all cases.}
\label{dmrem}
\end{figure}

Fig.\ \ref{dmrem} shows the mass-fraction of compact remnants for five individual cases, drawn from runs 
with both small and large numbers of stars. There 
are considerable differences in the mass-fractions of individual runs arising from differences in 
the dissolution times:
Small-$N$ clusters close to the galactic center dissolve considerably faster (in Myrs) 
and turn a much lower fraction of bound stars into white dwarfs than clusters at larger $R_G$. 
Except for the small-$N$ cluster close to the galactic center,
we always find that the mass fraction of compact remnants is near 90 \% close to the end of the cluster lifetime.
After we subtract the increase due to stellar evolution according to eq. \ref{fcrsev} from the different
cases, the differences become rather small (dashed lines). In agreement with Vesperini \& Heggie (1997), we find
that the mass-fraction of compact remnants is still 
increasing since low-mass main-sequence stars are preferentially lost from the clusters.
From Fig.\ \ref{dmrem}, we obtain the following fit to the overall evolution:
\begin{equation}
 F_{CR} = \left. F_{CR} \right|_{SEV} + 0.18 \; (T/T_{DISS})^2 + 0.16 \; (T/T_{DISS})^3  . 
\label{fcr}
\end{equation}

\begin{figure*}
\epsfxsize=17cm
\begin{center}
\epsffile{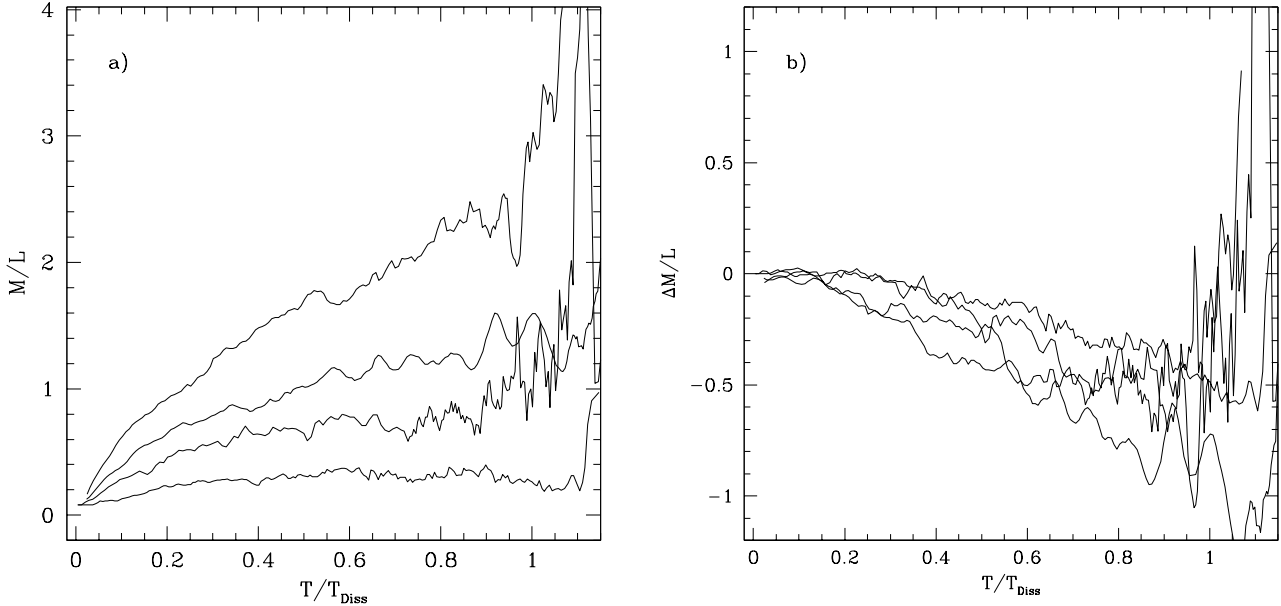}
\end{center}
\caption{Cluster mass-to-light ratios for the same clusters as in Fig.\ \ref{dmrem}. Panel a) shows
 the full evolution, panel b) shows the remaining part after the effect of stellar evolution has been subtracted.
 Dynamical evolution
 decreases the average $M/L$ ratio by about 0.3 to 0.7 at $T/T_{DISS}=0.8$ and increases it towards the end of the lifetime.}
\label{mlr}
\end{figure*}

A similar analysis can be done for the cluster mass-to-light ratios. Fig.\ \ref{sev} b) depicts the
evolution of the $M/L$ ratio for a population of stars which evolves due to stellar evolution alone.
Clusters start with an $M/L$ ratio of 0.08. This value increases constantly as the most luminous 
stars disappear.
After 13 Gyrs, the $M/L$ ratio has reached 2.0, which is within the range
of mass-to-light ratios observed for galactic globular clusters (Mandushev et al.\ 1991, Pryor \& Meylan 1993). 
For $T>100$ Myr, we 
can fit the evolution of the MLR by the following equation:
\begin{equation}
 \left. M/L \right|_{SEV} = 3.632  - 5.628 z + 3.364 z^2  - 0.919 z^3 + 0.1 z^4 \; , 
\label{mlrel}
\end{equation}
where $z$ is again $z = \log_{10}(T)$. The scatter around the mean relation is somewhat larger than before 
since the cluster luminosity is dominated by relatively few giants. 

Fig.\ \ref{mlr} a) shows the evolution of the $M/L$ ratios for four different clusters. In general, all mass-to-light
ratios are increasing, although with considerable differences between individual runs. Differences arise 
since clusters which dissolve quicker do so at an age in which they still contain
many luminous stars. Panel b) of Fig.\ \ref{mlr} shows the $M/L$ ratios for the same clusters after the effect of stellar evolution
due to eq.\ \ref{mlrel} has been subtracted: 
\begin{equation}
\Delta M/L = M/L - \left. M/L\right|_{SEV} \; .
\end{equation}
The remaining change is due to the dynamical evolution
alone. It can be seen that dynamical evolution leads to a decrease of the $M/L$ ratios because
faint, low-mass stars are preferentially lost from the clusters. On the 
other hand the fraction of compact remnants which contribute almost nothing to the cluster luminosity is increasing,
so that close to the end the $M/L$ ratios increase again.
For galactic globular clusters, we would predict that clusters in an advanced 
stage of their evolution have smaller mass-to-light ratios than clusters of similar age which have
lost only small part of their mass, the maximum differences being about 
$\Delta M/L$ = 0.5. Mandushev
et al.\ (1991) and Pryor \& Meylan (1993) have indeed found a correlation of the average $M/L$ ratio of a cluster with its
mass in the sense that low-mass clusters have smaller $M/L$ ratios. This could be
the result of the dynamical evolution since, for example, any cluster within a few kpc from the galactic center
 which has a current mass $M < 10^5 M_\odot$ 
must have lost most of its initial mass and should be close to final dissolution (see Fig.\ \ref{mimass}). 

\subsection{Mean masses and central remnant fraction}

\begin{figure}
\epsfxsize=8.3cm
\begin{center}
\epsffile{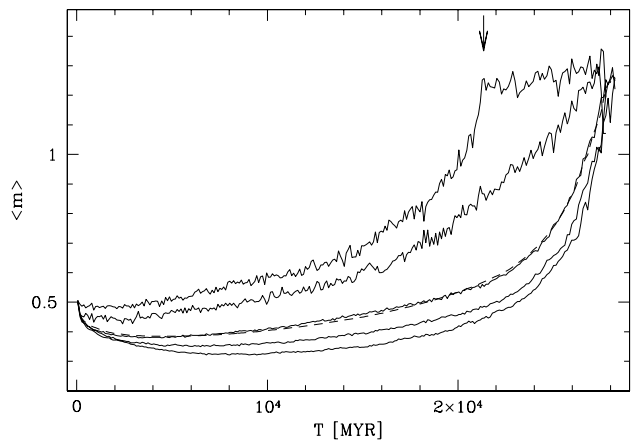}
\end{center}
\caption{Evolution of the mean mass of all stars
 with time for the run with $N=128$K stars, $W_0=5.0$ and $R_G = 8.5$ kpc.
   Shown are the mean mass of all stars in the cluster (dashed line) and (from top
 to bottom) mean masses in lagrangian shells containing 1\%, 5-10\%, 40-50\%, 70-80\% and 90-100\% of the bound mass. 
  The time of core collapse is marked by an arrow. 
 The mean mass of stars near the half-mass radius is always close to the mean mass of all stars   
  in the cluster.}
\label{meanm}
\end{figure}

Fig. \ref{meanm} depicts the evolution of the mean masses of all stars in our standard run.
In the beginning, mean masses drop throughout the cluster due to the mass-loss of the most massive stars from 
stellar evolution. For the inner radii, this is soon compensated by
mass segregation, which causes heavy-mass remnants and main-sequence stars to sink towards the cluster center. 
As a result, the mean mass in the core rises until core-collapse, when it reaches
a value of $<\!m\!> = 1.2 \; \mbox{M}_\odot$. This is similar to the masses of massive white dwarfs and neutron stars
and shows that in the center compact remnants must make up a significant fraction of the stellar population 
by this time.
After core-collapse, the mean mass in the center remains nearly constant.
Outside the half-mass radius, mean masses decrease initially as a result of stellar evolution
and the drift of massive stars into the core. Later, they increase
since low-mass stars are preferentially lost from the cluster. In the end, mean masses are almost
the same throughout the cluster. The mean mass around the
half-mass radius stays close to the mean mass of all stars in the cluster (dashed line).

\begin{figure}
\epsfxsize=8.3cm
\begin{center}
\epsffile{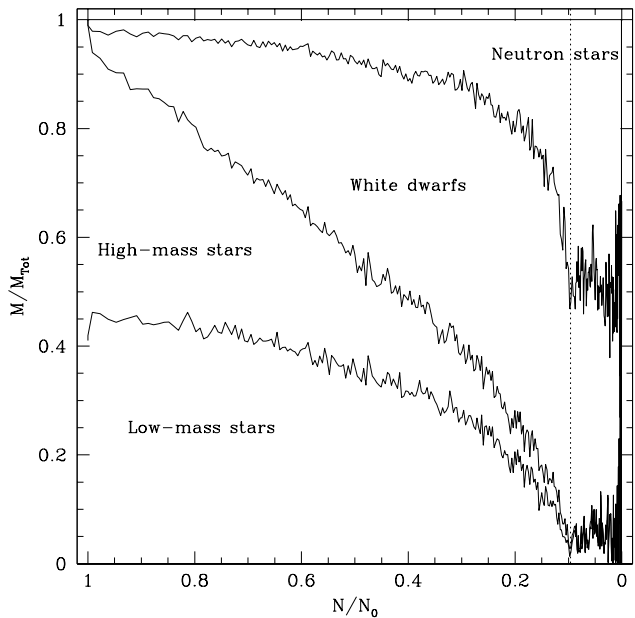}
\end{center}
\caption{Evolution of the central mass fraction of compact remnants
  for the same run as in Fig.\ \ref{meanm}. Since mass-segregation causes compact remnants
 to sink into the center, their mass fraction is increasing faster in the center 
  than in the rest of the cluster. By the time of core-collapse (dotted line),
  the center is almost entirely made up out of neutron stars and white dwarfs.}
\label{mdistc}
\end{figure}

Fig. \ref{mdistc} shows how 
the mass-fraction of white dwarfs and neutron stars changes in the core as a function of the number of stars left in
 the cluster. Here, the core was 
defined as the cluster part which contains the innermost 2\% of the mass. The mass-fraction of remnants
increases faster in the center
than in the rest of the cluster since white dwarfs and neutron stars sink towards the center due to mass-segregation.
By the time of core-collapse, the central remnant fraction has reached 95\%, despite the fact 
that their overall mass-fraction is only 60\% (see Fig.\ \ref{mdist}). At core-collapse 40\% of the central mass is in the form
of neutron stars. This value stays nearly constant in post-collapse, since most neutron stars have masses of
order 1.2 $M_\odot$, similar to the mass of the most massive white dwarfs. Hence, the relative mass-fraction of
both components should not be altered much by the dynamical evolution of the cluster. 

Most other clusters also show large 
central remnant fractions of order 90\% or more. The exact value
depends on the time at which core-collapse happens: Clusters which undergo core-collapse
later and have turned a larger fraction of their stars into compact remnants show generally also larger
central remnant fractions. 

Lugger et al.\ (1995) have measured U-band CCD surface brightness profiles for
15 globular clusters out of the sample of 21 clusters identified to be core collapse clusters by Djorgovski \& King (1986).
They found that 13 of them have power-law slopes in their centers with a mean slope of $\sigma(r) =r^{-0.85 \pm 0.1}$.

\begin{figure}
\epsfxsize=8.3cm
\begin{center}
\epsffile{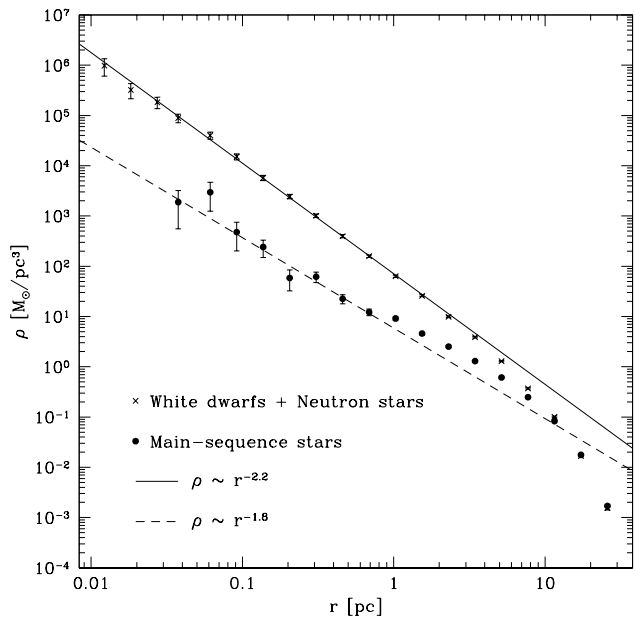}
\end{center}
\caption{Mass density at core collapse for the run with $\epsilon = 0.5$, $N=128$K stars.
 The combined mass density of white dwarfs and neutron stars is shown by crosses, the density profile of 
  main-sequence stars by dots. The central density of compact remnants exceeds that of main-sequence stars
  by more than a factor of 30 in the core.}
\label{tccdprof}
\end{figure}

Fig.\ \ref{tccdprof} shows the mass density profile of the $N=128$K star, $\epsilon=0.5$ run at core-collapse time.
We have chosen this cluster since it has a large number of stars remaining at core-collapse and also 
a core-collapse time which is close to the present time. It might therefore be
one of our best analogues to a large galactic globular cluster. The strong central concentration of
white dwarfs and neutron stars can clearly 
be seen. Inside 0.1 pc, their mass-density is more than a factor of 30 higher than that of main-sequence stars. The 
density
profile of compact remnants can be approximated by a power-law with slope $\rho \sim r^{-2.2}$ from the centre 
out to about 2 pc. This slope is in good agreement to what Takahashi et al.\ (1997) found from Fokker-Planck
simulations of multi-mass clusters without mass-loss. The main-sequence stars
follow a shallower profile then the compact remnants inside 0.5 pc with $\rho \sim  r^{-1.8}$. 
Outside this radius, their density profile decreases is even shallower up to 3 pc. Beyond this radius the profile 
steepens again due to the presence of the tidal field. Since the central density of main-sequence stars
is much smaller than that of the compact remnants, core-collapse is done mainly by the remnants and not by
the main-sequence stars. Projected and number density distributions show a similar behavior. For the
projected number densities we obtain slopes of $\sigma_{CR} \sim r^{-1.1}$ and  $\sigma_{MS} \sim r^{-0.85}$
for compact remnants and main-sequence stars respectively. 
Our density profile for the main-sequence 
stars is therefore in very good agreement to what Lugger et al.\ (1995) found for the mean projected density 
distribution. 
Furthermore, several clusters in their paper, e.g. NGC 6284, 6325, 6397 and 6522, also show a saddle like profile
at intermediate radii between the core and the halo. This could be an indication that in these clusters too
most mass in the centre is in the form of white dwarfs and neutron stars. In contrast, main-sequence stars in the 
run with $W_0=7.0$ and$N=128$K 
stars follow a power-law profile after core-collapse from the center all the way up to several pc. A discussion 
where the properties of this cluster are compared with the center of the globular cluster M15 and the implications
for a possible detection of a intermediate mass black hole in this cluster (Gerssen. et al.\ 2002) can be found in 
Baumgardt et al.\ (2002). 
 
A close inspection shows that most white dwarfs in the core tend to be old, high-mass white dwarfs which are
already very faint. In addition, the neutron stars, if not members in close binary systems, would also be 
difficult to detect. In principle,
a centrally concentrated distribution of faint stars could appear as an intermediate mass 
central black hole to observers, since they also give rise to an increasing
velocity dispersion of the main-sequence stars.
In order to test if the compact remnants in the core could appear as a black hole, we had a look
at the central mass-to-light ratio, since a large central mass-to-light ratio could be taken as an indication
for the presence of a massive black hole in the cluster center. 

\begin{figure}
\epsfxsize=8.3cm
\begin{center}
\epsffile{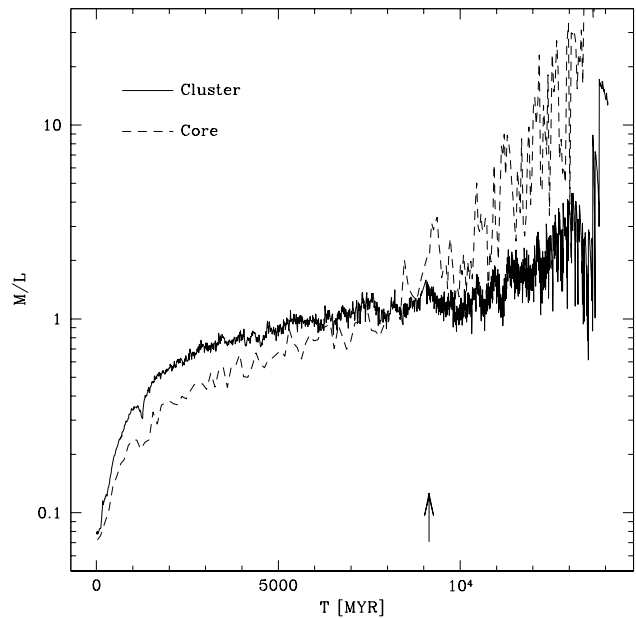}
\end{center}
\caption{Total and central mass-to-light ratios for the same run as in Fig.\ \ref{tccdprof}.
 Central mass-to-light ratios were calculated for the shell containing 2\% of all main-sequence stars
 in projection.
 The arrow marks the time of core-collapse. After core-collapse, central mass-to-light ratios can
 reach values of 10 or more due to the high fraction of compact remnants in the core.}
\label{mlrc}
\end{figure}

Fig.\ \ref{mlrc} shows the evolution of the central mass-to-light ratio for the
$N=128$K star, $\epsilon = 0.5$ run. Here we defined as core the radius of the shell which
contains the innermost 2\% of main-sequence stars in projection. 
Since high-mass stars sink preferentially into the core, the central mass-to-light ratio is in the beginning smaller 
in the core than in the rest of the cluster. This changes as more and more white dwarfs and neutron stars accumulate 
in the core, and after core-collapse the central mass-to-light ratio reaches values of 10 or more. This could
explain the high central $M/L$ ratios found for some globular clusters: D'Amico et al. (2002) for example found 
that an $M/L$ ratio of 10 is necessary to explain the pulsar timings in NGC 6752.
Since this cluster has a light-profile
similar to a post-collapse cluster (Djorgovski 1993), the large mass-to-light ratio could be due to a
concentration of compact remnants in the cluster core. 

A major uncertainty of the present analysis is that our clusters did not contain primordial binaries.
Since at least some fraction of these binaries would contain main-sequence stars with masses of $m \approx
0.7 M_\odot$, the total mass of the binary would be comparable to, or even exceed the masses of high-mass
white dwarfs, so these binaries could
also sink towards the core, thereby decreasing the central mass-to-light ratio. On the other hand, 
core-collapse can be achieved only if the binary fraction is decreased considerably from the initial value
so that there is no significant energy release from binaries any more. In addition, any main-sequence star
in a binary which has reached the center is likely to be replaced by a more massive compact remnant through
dynamical interactions, and would subsequently be expelled from the core. 
Central mass-to-light ratios might therefore be similar to the present values even if clusters started with
a significant fraction of primordial binaries. 

\subsection{Energy equipartition}

Any self-gravitating system which contains stars of different masses will show a tendency towards
energy partition, such that at any radius heavy stars have the same kinetic energies as lighter ones. 
If stars start with the same average velocities initially, this implies that heavy stars sink towards 
the center and lighter stars will move
outwards. If the density of high-mass stars in the cluster center becomes too large, they will
form a self-gravitating system of their own, in which case equipartition of energies cannot be 
reached (Spitzer 1969).

In order to test for equipartition of energies in our runs, we define 
a correlation parameter between the energy of a star and its mass as
\begin{equation}
 P = \frac{\sum_i \; (<\!T\!>-T_i) \; (<\log{m}>- \log{m_i})}{\sqrt{\sum_i (<\!T\!>-T_i)^2} \; \sqrt{\sum_i (<\log{m}>- \log{m_i})^2}}
\end{equation}
where the sums run over all stars within a radial shell and $<\!T\!>$ and $<\!\log{m}\!>$ are the mean kinetic 
energy and logarithm 
of the mass of stars in the shell. The logarithm of the mass was chosen in order to give less weight 
to the few stars at the high-mass end. $P$ will be zero if kinetic energy is uncorrelated with mass, i.e.
energy equipartition is reached. 
\begin{figure}
\epsfxsize=8.3cm
\begin{center}
\epsffile{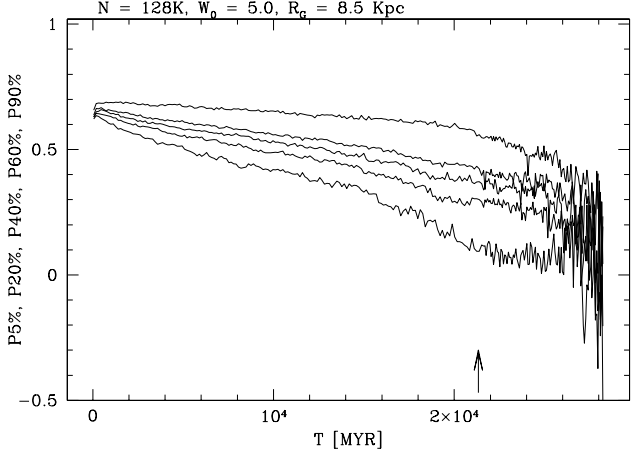}
\end{center}
\caption{Time evolution of the energy equipartition parameters $P$ for 5 different lagrangian shells for the
 run with $N=128$K stars, $W_0=5.0$ and $R_G=8.5$ kpc. Shown are parameters for the shells containing (from bottom to top)
  5\%, 10-20\%, 30-40\% 50-60\% and 80-90\% of the mass. Energy
  equipartition is reached at the center at core-collapse time (marked by an arrow). Outside the half-mass radius,
  it is not reached within the lifetime of the cluster.}
\label{bpar1}
\end{figure}

\begin{figure}
\epsfxsize=8.3cm
\begin{center}
\epsffile{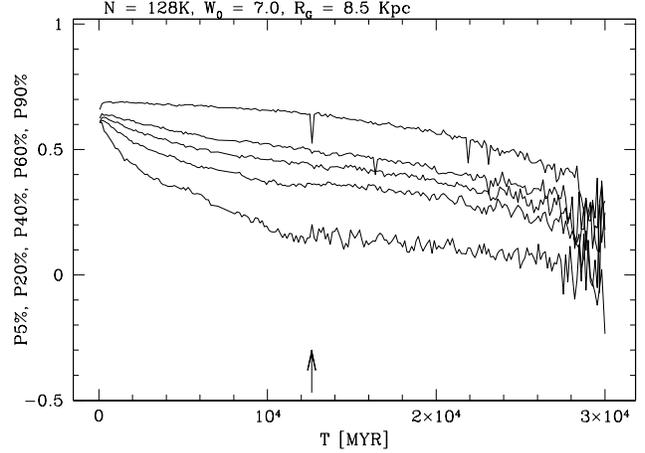}
\end{center}
\caption{Same as Fig.\ \ref{bpar1} but now for the cluster with initial concentration 
 $W_0 = 7.0$ and $N=128$K stars.}
\label{bpar2}
\end{figure}

Figs.\ \ref{bpar1} and \ref{bpar2} show the evolution of the $P$ parameters for 5
different lagrangian shells for runs with initial central concentrations of $W_0=5.0$ and $W_0=7.0$ respectively. 
The arrows mark the time of the first core-collapse in both runs.
Energy equipartition is reached for the inner shells at around core-collapse time. For radii 
near the half-mass radius,
energy equipartition is reached only close to the end of the lifetime and for the halo it is never reached. This is the
case even in the run with $W_0=7.0$, despite the fact that this run spends nearly half its lifetime in the post-collapse phase.
We find this result for all our runs and
similar results were also obtained by Giersz \& Heggie (1996) in their simulations of
small-$N$ clusters without stellar evolution. We conclude that for globular clusters, energy 
equipartition, if not present in some way already initially,
can be expected to exist only in the centers of clusters which
went through a core collapse phase. 

\subsection{Velocity anisotropy}

It has long been known that isolated star clusters become radially anisotropic after core-collapse due to 
the scattering of stars out of the cluster core (Spitzer 1987). In a tidal field, the situation is complicated
by the outflow of stars over the tidal boundary. Giersz \& Heggie (1997) found in $N$-body simulations
that star clusters in tidal fields remain isotropic during their evolution except for the outer parts which become 
tangentially anisotropic due to the preferential loss of stars on radial orbits. Similar results were also
found by Takahashi, Lee \& Inagaki (1997) and Takahashi \& Lee (2000) in their anisotropic
Fokker-Planck simulations. 
There is also mounting observational evidence that at least some globular clusters, like e.g. $\omega$ Cen (van
Leeuwen et al.\ 2000), have anisotropic velocity distributions, although in this case it is not clear 
whether such velocity distributions were primordial, or formed as a result of the dynamical
evolution.

\begin{figure}
\epsfxsize=8.3cm
\begin{center}
\epsffile{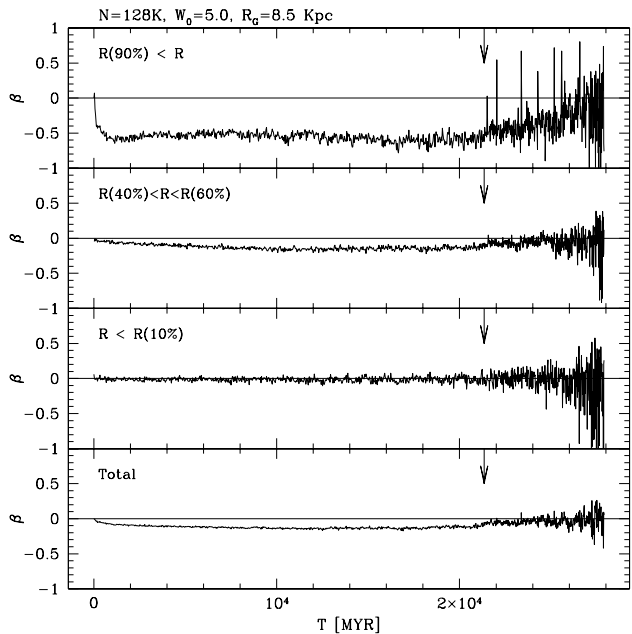}
\end{center}
\caption{Evolution of the velocity anisotropy with time for different radii for the run with $N=128$K stars,
 $W_0=5.0$ and $R_G=8.5$ kpc. Arrows mark the core-collapse time. Tangential 
  anisotropy develops rapidly in the outer parts. Its size remains more or less constant until half-mass time, 
  after which it decreases slowly. In the inner parts, clusters stay 
  isotropic throughout the evolution.} 
\label{aniso1}
\end{figure}

Since it is difficult to incorporate all effects of an external tidal field into a 
Fokker-Planck code and since most $N$-body simulations performed contain only relatively few stars, it might
be interesting to look for signs of velocity anisotropies in our runs too.
Following Binney \& Tremaine (1986), we define an anisotropy parameter $\beta$ as
\begin{equation}
\beta = 1 - \frac{v^2_t}{2 \; v_r^2} \;\; ,
\end{equation}
where $v_r$ and $v_t$ are the radial and tangential velocity dispersions. They are measured
in a non-rotating coordinate system in which the cluster center is always at rest at the origin. 
Fig. \ref{aniso1} shows the evolution of the velocity anisotropy for the run with $N=128$K stars from our
standard family.
It can be seen that the outer parts become rapidly tangentially anisotropic. Inspection of the orbits shows that most
stars in the outer parts are counterrotating, probably due to the fact that retrograde orbits
are more stable against escape than prograde orbits (Keenan \& Innanen 1975). 
As a consequence, the clusters also start to rotate in their outer parts. Fig.\ \ref{aniso1} shows
that the tangentially anisotropic velocity dispersion is restricted to the outermost radii. 
We find a similar behavior for the counterrotation of the stars. 
We note that the amount of tangential velocity anisotropy and rotation in the outer parts would have been larger
if, as is frequently done, we would have chosen a rotating coordinate system as reference in which 
the direction from the cluster to the galactic center remains fixed.

\begin{figure}
\epsfxsize=8.3cm
\begin{center}
\epsffile{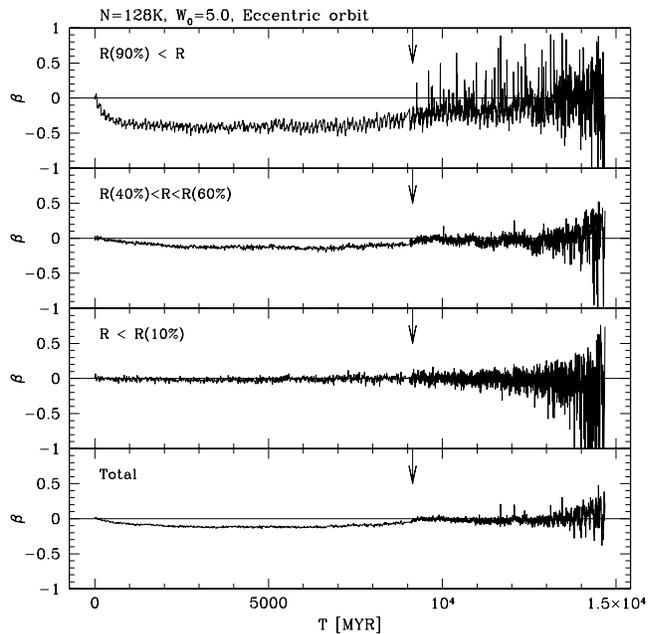}
\end{center}
\caption{Same as Fig.\ \ref{aniso1} for a cluster in an eccentric orbit with $\epsilon=0.5$. Tangentially
  anisotropic velocity distributions still form in the outer parts, but the amount of anisotropy is 
  reduced.}
\label{aniso2}
\end{figure}

The amount of tangential anisotropy remains more or less constant until near core-collapse, when it starts 
to decrease. An explanation could be that the relaxation time of a cluster decreases as it loses mass and
that a smaller relaxation time in the halo implies a more efficient isotropization of the stellar orbits. 
In addition, more stars are scattered out of the core into the outer cluster areas in the post-collapse phase. 
Since these stars will be on near radial orbits initially,
they will also decrease the amount of tangential anisotropy in the halo. 
For clusters in eccentric orbits we obtain similar results, except that the
amount of tangential anisotropy is smaller compared to the circular case (Fig.\ \ref{aniso2}) Globular cluster
might therefore also have slightly tangentially anisotropic velocity distributions in their halos.

Except for the outermost
radii, clusters are nearly isotropic and there is no sign of the radially anisotropic velocity
dispersion which forms in isolated clusters. These results 
confirm earlier results obtained by Takahashi \& Lee (2000) and Giersz \& Heggie (1997).

\begin{figure}
\epsfxsize=8.3cm
\begin{center}
\epsffile{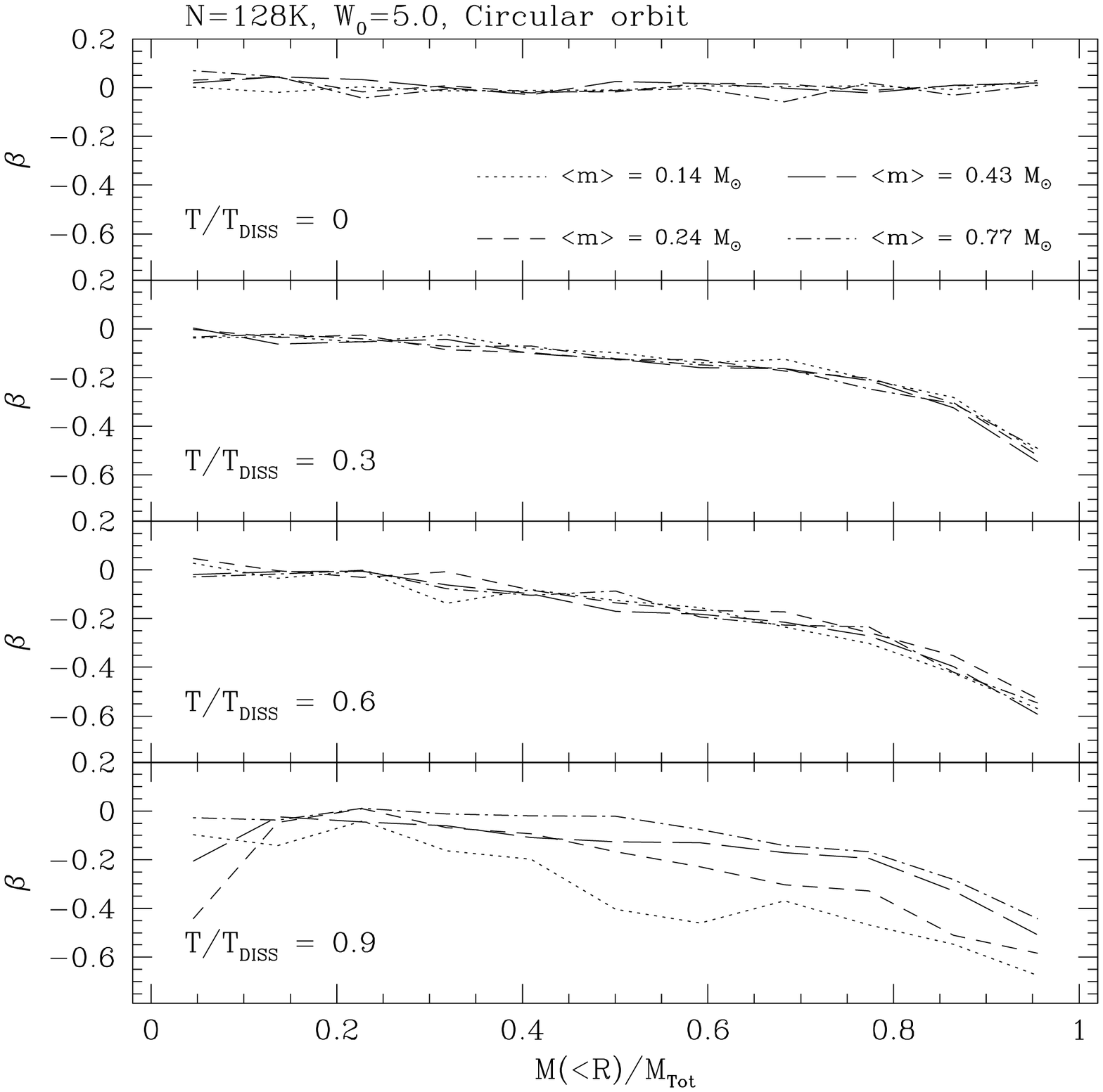}
\end{center}
\caption{Velocity anisotropy for different mass groups as a function of enclosed mass
  for the same run as in Fig.\ \ref{aniso1}. The differences remain
  small until close to final dissolution, when low-mass stars have more tangentially
 anisotropic distributions by up to $\beta = 0.3$ in the outer parts.} 
\label{aniso3}
\end{figure}

Takahashi \& Lee (2000) found that there is a difference in the mean anisotropies of high and low-mass stars,
especially in the cluster halos. In their simulations, high-mass stars always followed more radially anisotropic 
velocity distributions than low-mass stars, the difference depending on the fraction of mass lost from a cluster 
and the distance from the cluster center. 
Fig.\ \ref{aniso3} shows the anisotropy parameter $\beta$ for 4 different groups of main-sequence stars  
at several times during the cluster evolution. Close to the end,
the velocity distributions become indeed different. In the halo we find differences of up to $\beta=0.3$,
while near the cluster center differences remain small at any time.
The differences could be due to 
mass-segregation: Since massive stars are more centrally concentrated, they tend to be located at lower 
energy levels which have more isotropic velocity distributions. Since mass-segregation itself develops 
only fairly late in the halo, it also takes very long until there are differences in the anisotropies.

\section{Comparison with galactic globular clusters}   

\subsection{Dissolution times and the depletion of the galactic globular cluster system}

Dinescu et al.\ (1999) have compiled a catalog of absolute proper motions for 38 globular clusters.
Besides the cluster orbits, they also determined destruction rates for individual
clusters, following the formalism of Gnedin \& Ostriker (1997). Their lifetimes provide a good
check for our formula for the lifetime. In order to compare the lifetimes, we first took the orbital 
parameters for each cluster from Table 5 of Dinescu et al.\ (1999) and calculated the present day cluster masses
$M_C$ from the absolute magnitudes in the McMaster database of globular cluster parameters (Harris 1996), assuming 
a mass-to-light ratio of $M/L_V = 2$.  
We then calculated iteratively the initial mass $M_0$ necessary to produce a cluster with 
mass $M_C$ after a Hubble time and the remaining lifetime for each cluster, using eqs.\ \ref{gtime} and \ref{emfunc}.
\begin{figure*}
\epsfxsize=17cm
\begin{center}
\epsffile{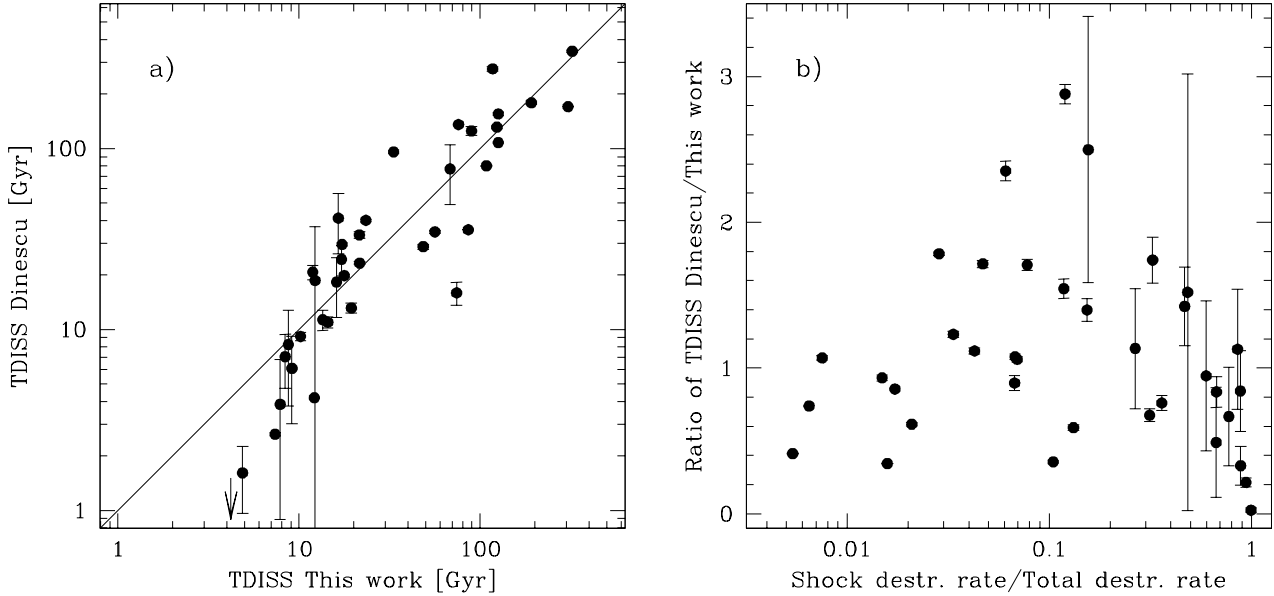}
\end{center}
\caption{a) Lifetimes from Dinescu et al.\ (1999) against lifetimes derived in this work for a
  sample of 38 globular clusters. There is good agreement except for clusters
  close to dissolution for which we predict slightly longer lifetimes. b) Ratio of the lifetimes
  as a function of the relative importance of tidal shocking compared to the total destruction rate
  as given by Dinescu et al. (1999). For clusters for which shocking is important, Dinescu et al. (1999) derive
  smaller lifetimes.}
\label{dindiss}
\end{figure*}

Fig.\ \ref{dindiss} compares our results for the remaining lifetime with the destruction rates obtained by
Dinescu et al.\ (1999) for the galactic model of Paczy\'nski (1990). There is quite good agreement 
between the lifetimes determined from $N$-body simulations and those determined from Fokker-Planck
calculations. The exception might be clusters close to destruction, where the lifetimes of Dinescu 
et al.\ (1999) seem to be somewhat smaller than ours. Panel b) compares the differences in the lifetimes 
as a function of the importance of disc and bulge shocking for the total destruction rate. Both parameters 
are taken from Dinescu et al.\ (1999). It can first be seen that the clusters for which tidal shocks
are least important have longer lifetimes according to our formula. Close inspection of the orbits shows that
nearly all of these clusters are orbiting at galactocentric distances larger than 10 kpc, so the differences 
could be due to differences in the underlying galactic models. Secondly, clusters which dissolve mostly due to
external tidal shocks also have longer lifetimes according to our formula. This could be a hint that our neglect 
of disc-shocking overpredicts the lifetimes in these cases. However, since most of these 
clusters are also close to dissolution,
it is difficult to determine what is responsible for the difference. Nevertheless, the overall agreement is 
reasonably good and we find the same agreement for the lifetimes obtained by Dinescu et al. for a 
different galactic model.

We finally note that the lifetimes of eq.\ \ref{gtime} are also in fair agreement with those
obtained by Vesperini \& Heggie (1997, eq.\ 11) and Giersz (2001, section 3.2) for clusters with
masses around $M_C=10^3$ to $10^4 M_\odot$, despite the fact that the initial mass-functions used 
and the way stellar evolution is modeled differ slightly. For larger clusters, we
predict lifetimes which are smaller by a
factor of 2 to 5 since eq.\ \ref{gtime} leads to a scaling of the cluster lifetime with cluster mass flatter 
than a scaling with the relaxation time.

\begin{figure}
\epsfxsize=8.3cm
\begin{center}
\epsffile{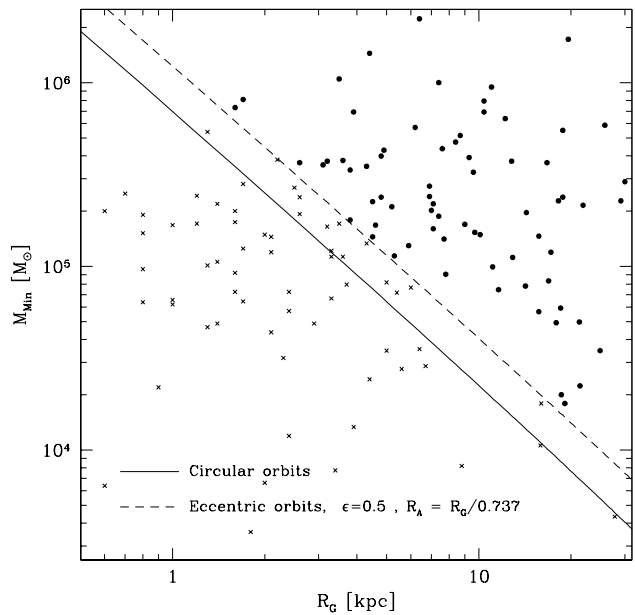}
\end{center}
\caption{Galactocentric distance vs. mass of galactic globular clusters. The solid line shows the minimum
  mass
  needed such that a cluster survives for another Hubble time, assuming it moves in a circular orbit. The dashed
  line shows the same for clusters in eccentric orbits with $\epsilon=0.5$ if one assumes that each cluster is
  located at the mean galactocentric
   distance of such an orbit $<\!\!R_G\!\!> = 0.74 R_A$. Clusters which will be destroyed in the eccentric case
  are shown by crosses, clusters which survive by dots. In the eccentric case, 67\% of all globular
  clusters will be destroyed within the next Hubble time.}
\label{mimass}
\end{figure}

As an application of our results for the lifetimes, we will have a look at how dynamical evolution depletes
the galactic globular cluster system.
Fig.\ \ref{mimass} compares masses and distances of galactic globular clusters with the minimum
mass needed such
that a cluster survives for another Hubble time. The distances and masses are again taken from Harris (1996).
The minimum mass was determined by first using eq.\ \ref{gtime} to
calculate the mass of a cluster which survives for two Hubble-times and then eq.\ \ref{emfunc} to
determine its present day mass. We calculated lifetimes by assuming that all clusters either move on
circular or eccentric
orbits with $\epsilon = 0.5$. For the eccentric orbits we used the fact that the ratio of the
time-averaged galactocentric
distance of a cluster in an $\epsilon=0.5$ orbit to its apogalactic distance is equal to
$<\!\!R_G\!\!>/R_A = 0.74$ if the cluster is moving through an logarithmic galactic potential.
All present distances are divided by this ratio to obtain an estimate of $R_A$ for each
cluster.

It can be seen that a large fraction of galactic globular clusters will not survive for another Hubble
time.
Depending on which orbital type we assume, we find that between 53\% (circular) to 67\% (eccentric orbit)
of all
galactic globular clusters will be destroyed, most of them in the inner parts of the galaxy.
The higher destruction
rate of eccentric orbits is probably closer to the truth since the galactic globular cluster system
as a whole has a radially anisotropic velocity distribution (Dinescu et al. 1999), implying 
rather large orbital eccentricities in the mean.

Depending on which orbital type we assume, we find that currently between 7 to 10 globular clusters are 
destroyed per Gyr, which is in agreement 
with the value of $5 \pm 3$ found by Hut \& Djorgovski (1992) from observational estimates 
of the relaxation time in galactic globular clusters.
Similarly, Fall \& Zhang (2001) found by semi-analytical modeling that at present between 5 to 8\% of 
all globular clusters are destroyed per Gyr, which corresponds to 7 to 11 clusters. If we 
assume an exponential decrease with time, this would lead to destruction rates between 50\% to 67\% after
13 Gyr. Hence, although the assumptions about the underlying kinematic and galactic model are quite
different in each paper, there is good agreement with our values. 

\subsection{Evolution of the mass-function}

In this section we will apply the results of previous chapters to a sample of galactic globular clusters
with well observed parameters.
In recent years, the proper motions of a number of clusters could be measured and tied to the reference
frame of the Hipparcos Catalogue so that their orbits are relatively accurately known (see Odenkirchen 
et al. 1997, Dinescu et al.\ 1999 and references therein). 
In addition, accurate determinations of the mass-functions at the low-mass end were obtained with the 
HST for about a dozen globular clusters (Piotto \& Zoccali 1999, Paresce \& de Marchi 2000
and references therein). 

Table 2 lists 10 clusters for which orbital information and determinations of the slope
of the mass-functions are simultaneously available. Beside the names, we have listed the present-day
masses, calculated from the absolute luminosities given in the compilation of Harris (1996)
and an assumed mass-to-light ratio of 2. The perigalactic and
apogalactic distances of the clusters in columns 3 and 4 were taken from Dinescu et al.\ (1999), 
except for Pal 5, 
where the data for the mass and the orbital information are from Odenkirchen et al.\ (2001, 2002). 
Column 5 lists the 
observed slope of the mass-function at the low-mass end, determined mostly from observations taken around
the half-light radius. 

Column 6 lists estimated slopes for the global mass-function, determined 
by correcting the local mass-function for mass-segregation with the help of multi-mass King-Michie
models. Since this method assumes energy partition throughout the cluster, and since we have seen that 
energy equipartition is not established in the center until core-collapse, these corrections could 
be in error for some clusters. However, as can be seen in Table 2, the corrections
made are usually fairly small and should therefore not significantly influence our conclusions.
The values in columns 5 and 6
come from Piotto \& Zoccali (1999), except for Pal 5 and NGC 6121, where we have taken data from Grillmair \&
Smith (2001) and Bedin et al.\ (2001) respectively. For both clusters, observations were
done around the half-mass radius. In addition, Grillmair \& Smith (2001) did not find
strong evidence for mass-segregation in case of Pal 5. Hence we assume for both clusters that the
global mass-functions have the same slopes as the measured ones. 

\input{table2.tex}

In order to make predictions for individual clusters, we will neglect the influence of a galactic disc and
assume that clusters move through a spherical logarithmic potential. 
We then proceed similar to the previous section and calculate for each globular cluster the eccentricity 
of its orbit from columns 3 and 4 of Table 2 and estimate its dissolution time and its initial mass
by using eqs.\ \ref{gtime} and \ref{emfunc}. These values can be found in columns 7 and 8
of Table 2. 

The cluster Pal 5 has the smallest lifetime in the sample according to our predictions. 
This is also confirmed by observations, since Odenkirchen et al.\ (2001, 2002) find evidence that this cluster has
strongly populated tidal tails and a very low mass, which both indicate that it must be 
close to final dissolution.
 
From a comparison of the dissolution time with the cluster ages,
the global slope of the mass-function of low-mass stars and
the dark-matter content were finally determined by using eqs.\ \ref{mftf}, \ref{fcrsev} and \ref{fcr}.
These values can 
be found in the last two columns of Table 2. Since our simulations do not consider the effect of disc-shocking,
our dissolution times are likely to be upper limits and the changes that the mass-functions underwent
could on average be larger than estimated by us, especially for clusters which are close to final dissolution.
The differences might however not be very large: Gnedin \& Ostriker (1997) for example
found that the mean destruction rate is increased by about 30\% compared to the
pure relaxation case if their simulations include a galactic disc. According to their simulations, 
the influence of a disc is strongest for a cluster near the galactic center and decreases further outward. 
At $R_G = 3$ kpc for example, 
which corresponds roughly to the
median perigalactic distance of the clusters in Table 2, the inclusion of a disc increases the average 
destruction rate by only 10\%. Similarly, the data listed by Vesperini \& Heggie (1997) in their Table 2
indicates that the influence of disc-shocking is diminishing with increasing distance from the galactic center
and seems to enhance the mass-loss rate at distances of several kpc by no more than 10\%. An inclusion
of a galactic disc would therefore probably change our results by only a small amount. 
 
\begin{figure*}
\epsfxsize=17cm
\begin{center}
\epsffile{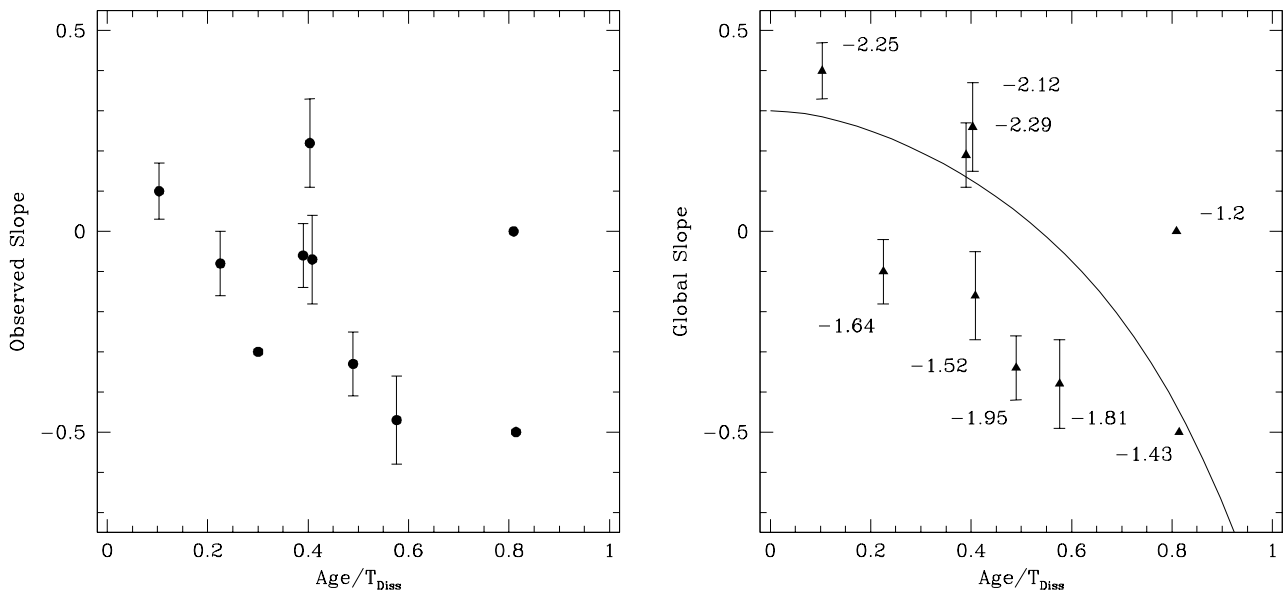}
\end{center}
\caption{Slopes of the mass-function against the fraction of time spent until complete dissolution for 
the clusters
 of Table 2. Observed mass-functions are shown in the left panel, global mass-functions are shown
 in the right panel. Errorbars are taken from Piotto \& Zoccali (1999) and show the uncertainties of the
  observed slopes. The solid line in the
 right panel shows our theoretical prediction.
 There is a clear correlation between
 the local and global slope of the mass-function and the dissolutionary stage of the cluster.}
\label{ltim1}
\end{figure*}

Fig.\ \ref{ltim1} depicts the observed and global slopes of the stellar mass-function for the clusters
in Table 2 in dependence of the fraction of 
time spent until complete cluster dissolution. Our simulations predict that the slope 
should decrease 
constantly as the clusters evolve, and there is indeed a good correlation between 
the slopes and the cluster lifetimes in both panels.
The only discrepant cluster is NGC 6121, which is close to final dissolution but 
still has a flat mass-function with slope $x = 0$. Since this cluster
has the orbit with the largest eccentricity
of all clusters in the sample, its small dissolution time is due mainly to its large orbital
eccentricity and could be significantly larger if the true orbit is less eccentric.
We also note that Richer et al.\ (2002) have determined a mass-function slope for NGC 6121
which corresponds to $x=-0.25$ in our notation. Using their value would remove most of the
discrepancy with our theory.
 
A Spearman rank-order correlation coefficient test shows that the observed and global slopes are 
correlated with the cluster lifetimes by confidence levels of 82\% and 97\%. Although one
should be cautious since the sample size is still rather small, we conclude that dynamical
evolution has depleted the stellar mass-function in globular clusters. 

The solid line in the right panel shows our prediction for the global mass-function. It is for all
dissolutionary stages within the range of observations, so it seems quite possible that the 
galactic globular cluster system has on average started with a slope of $x = 0.3$ at the low-mass
end, similar to what Zoccali et al.\ (2000) and Kroupa (2001) have found for the mean
mass-function of low-mass stars in the galactic bulge and disc.
At any dissolutionary stage, the scatter among different clusters is considerable 
and seems to exceed the errorbars, which would hint to additional parameters which influence the
mass-function. Next to binaries, metallicity effects have long been thought of influencing the mass-function
(McClure et al. 1986).
We have therefore plotted the metallicity from the McMaster database next to each cluster in the right panel.
It can be seen that all clusters with negative global slopes have small
metallicities while clusters with metallicities larger than $[Fe/H]=-2$ have slopes around $x= 0.3$. 
A second parameter which influences the stellar mass-function in globular clusters might therefore be 
the cluster metallicity. 
However, given the small number of data points and the fact that 
the mass-functions were deduced by different authors using various methods,
this conclusion is still uncertain.

\section{Conclusions}

We have performed a set of large scale $N$-body simulations of multi-mass star clusters 
moving through an external galaxy with a logarithmic density profile. 
Our simulations included a realistic spectrum of stellar masses, mass-loss due to stellar evolution, two-body 
relaxation and a fully consistent treatment of the external tidal field. The simulations were carried out
with the collisional $N$-body code NBODY4 on the GRAPE6 boards of Tokyo University and thanks to the speed 
of these boards we could perform simulations with particle numbers far greater than was possible hitherto. 
Our largest clusters have lifetimes well in excess of a Hubble time and become comparable to small
globular clusters in the later stages of their evolution.

Our results can be summarized as follows: 
All simulated clusters dissolve mainly as a result of two-body relaxation and external tidal truncation, with
the lifetimes of clusters moving on the same orbit scaling proportional to the relaxation time to a
power $x$ that is found to be in the range 0.75 to 0.85. This is in agreement with the 
results of Baumgardt (2001) for the scaling of the lifetimes of single-mass clusters in tidal fields. 
Clusters with stronger initial concentration $W_0$ show slightly larger exponents than less concentrated 
clusters. 
The scaling
law does not change significantly if one goes from the circular case to clusters moving on eccentric orbits,
which is a hint that relaxation is the dominant dissolution mechanism for eccentric orbits as well.
Utilizing our results for the lifetimes, we predict 
that about 2/3 of all galactic globular clusters will be destroyed within the next Hubble time.
This value is in good agreement with recent literature values and
argues for a strong evolution of the galactic globular cluster system. 

In agreement with Vesperini \& Heggie (1997), Takahashi \& Lee
(2000) and what one would
expect from the effect of mass-segregation, we find that low-mass stars are 
preferentially depleted from star clusters.
The depletion is strong enough to turn an initially increasing mass-function into one which is decreasing
towards low-mass stars. We find that the details of this process are nearly independent of the starting 
condition like cluster orbit, number of cluster stars or central concentration, and can be characterised
by a single parameter, as e.g. the fraction of time remaining until complete cluster dissolution or the
fraction of mass already lost from a star cluster.

Mass segregation causes mass-functions to be steeper in the inner parts than near the tidal radius
and might be difficult to correct for observationally since energy equipartition is established only very
late in the cluster evolution. In the center it is not complete until clusters have gone 
into core-collapse 
while in the outer parts it is not reached until clusters are almost completely dissolved. This confirms
an earlier result obtained by Giersz \& Heggie (1997) and shows that the application of multi-mass
King-Michie models to determine the global mass-function from observed local ones is not without risk.
A better strategy might be to conduct observations at or slightly beyond
the (projected) half-light radius, where we always find a good agreement between the observed slope of the
mass-function and the global one, and make no correction for mass-segregation at all.

The mass and number fractions of white dwarfs and neutron stars are increasing throughout cluster evolution
since these stars are the most massive components and are therefore least likely to escape.
For globular clusters near the
end of their lifetime, like e.q. Pal 5, we predict that most part of the cluster mass should be in the 
form of compact remnants, unless the initial mass-function had a slope significantly steeper than 
Salpeter above $m = 1 \; \mbox{M}_\odot$. White dwarfs and neutron stars are strongly concentrated towards
the center and central mass-to-light ratios can reach values of 10 or more at core-collapse time and
in post-collapse evolution. This could explain the recent detection of high central mass-to-light ratios in 
post-collapse globular clusters without the need to assume intermediate mass black holes in their centers
(Baumgardt et al. 2002). However, more detailed simulations with a larger number of cluster stars and 
containing a significant fraction of primordial binaries, possible for example with Monte-Carlo codes 
(Giersz 2001, Spurzem \& Giersz 2002), might be necessary to study this question in detail.

Clusters remain isotropic in their centers and near the half-mass radius and have mildly tangential anisotropies 
in their outer parts. The amount of tangential anisotropy depends on the orbit of the cluster and is largest
for clusters moving in circular orbits.
High-mass stars have slightly less tangentially anisotropic velocity distributions in the outer cluster parts 
although the differences remain small until close to the end.
In the outer parts, the simulated clusters also show a small amount of rotation since stars
orbiting in the same sense as the cluster moves around the galaxy are more easily lost from the cluster than stars 
in retrograde motion.

For a sample of galactic globular clusters with well observed parameters, we find a correlation between the 
observed slope of the mass-function
and the lifetimes predicted by us in the sense that clusters close to dissolution are depleted in low-mass
stars. There is also good agreement between our prediction for the slope of the mass-function at any evolutionary 
stage and the mean observed slope. 
It therefore seems conceivable that galactic globular clusters started on average
with slopes around $x = 0.3$ for $m<0.7 \; \mbox{M}_\odot$, which is similar to what was found by 
Zoccali et al.\ (2000) and Kroupa (2001) for low-mass stars in the galactic 
bulge and disc. This would indicate that no dramatic change has happened to the mass-function of low-mass stars within 
the last 10 to 12 Gyrs and that the mass-function does not depend much on the density of the environment where stars 
are formed. There are hints for additional parameters which influence the mass-function, as e.g. the cluster 
metallicity. However, given the fact that the number of globular clusters with well observed mass-functions is 
still rather small, this conclusion is uncertain. 
 
\section*{Acknowledgments}
We are grateful to Sverre Aarseth for making the NBODY4 code available to us and his constant help with the
program. We also thank Manuela Zoccali and Dana Dinescu for kindly sending us their data on globular clusters
and an anonymous referee for comments which helped improving the presentation of the paper.
H.B. is supported by the Japan Society for the Promotion of Science through Grant-in-Aid for JSPS 
fellows 13-01753.

\bsp
\label{lastpage}

\end{document}

%% file: table1.tex
\begin{table*}
\caption[]{Details of the performed $N$-body runs.}
\begin{tabular}{crccccrccr@{\hspace{0.1cm}}c@{\hspace{0.1cm}}rr@{\hspace{0.1cm}}c@{\hspace{0.1cm}}r}
\noalign{\smallskip}
Orbital & \multicolumn{1}{c}{$N$}& \multicolumn{1}{c}{$N_{Sim}$} &
\multicolumn{1}{c}{$W_0$} & \multicolumn{1}{c}{Orbit} & \multicolumn{1}{c}{$R_G^1$} & 
\multicolumn{1}{c}{$M_0$} & \multicolumn{1}{c}{$r_t$} &  \multicolumn{1}{c}{$\Delta M_{SEV}$} 
 & \multicolumn{3}{c}{$T_{Diss}$} & \multicolumn{3}{c}{$T_{CC}$} \\
Family & & & & & [pc] & \multicolumn{1}{c}{[$M_\odot$]} & [pc] & [\%] & \multicolumn{3}{c}{[MYR]} & \multicolumn{3}{c}{[MYR]}\\ 
\noalign{\smallskip}
(I)  &  8192 &  4 & 5.0 & Cir & 8500 &  4497.1 & 24.34 & 29.5  &  4149 & $\pm$ & 69 & 3329 & $\pm$ & 50\\
     & 16384 &  2 & 5.0 & Cir & 8500 &  9003.3 & 30.69 & 30.0  &  6348 & $\pm$ &135 & 5062 & $\pm$ & 66\\
     & 32768 &  1 & 5.0 & Cir & 8500 & 18407.6 & 38.95 & 32.1  &  9696 &       &    & 8412 & & \\
     & 65536 &  1 & 5.0 & Cir & 8500 & 36223.7 & 48.81 & 32.1  & 15197 &       &    &13193 & & \\
     &131072 &  1 & 5.0 & Cir & 8500 & 71236.4 & 61.15 & 32.1  & 23769 &       &    &21339 & & \\
\noalign{\smallskip}
(II) &  8192 &  4 & 7.0 & Cir & 8500 &  4401.6 & 24.17 & 27.8  &  3688 & $\pm$ & 75 & 1672 & $\pm$ & 42\\
     & 16384 &  2 & 7.0 & Cir & 8500 &  8927.7 & 30.60 & 29.8  &  5896 & $\pm$ & 65 & 2886 & $\pm$ &179\\
     & 32768 &  1 & 7.0 & Cir & 8500 & 18013.5 & 38.67 & 31.1  &  9694 &       &    & 4874 & & \\
     & 65536 &  1 & 7.0 & Cir & 8500 & 35611.1 & 48.53 & 31.9  & 15129 &       &    & 7869 & & \\
     &131072 &  1 & 7.0 & Cir & 8500 & 71699.2 & 61.28 & 32.9  & 25506 &       &    &12620 & & \\  
\noalign{\smallskip}
(III)&  8192 &  4 & 5.0 & 0.5 & 8500 &  4480.1 & 11.69 & 26.5  &  1961 & $\pm$ & 45 & 1007 & $\pm$ & 96\\
     & 16384 &  2 & 5.0 & 0.5 & 8500 &  9021.6 & 14.76 & 29.0  &  3142 & $\pm$ & 84 & 2194 & $\pm$ & 48\\
     & 32768 &  1 & 5.0 & 0.5 & 8500 & 17966.3 & 18.57 & 29.9  &  4899 &       &    & 3613 & & \\
     & 65536 &  1 & 5.0 & 0.5 & 8500 & 36156.8 & 23.45 & 31.2  &  7536 &       &    & 5924 & & \\
     &131072 &  1 & 5.0 & 0.5 & 8500 & 71385.2 & 29.42 & 31.5  & 11675 &       &    & 9332 & & \\
\noalign{\smallskip}
(IV) &  8192 &  4 & 5.0 & Cir & 2833 &  4441.8 & 11.66 & 23.5  & 1135 & $\pm$ & 38 &  853 & $\pm$ & 45\\
     & 16384 &  2 & 5.0 & Cir & 2833 &  9023.8 & 14.76 & 26.0  & 1892 & $\pm$ & 70 & 1576 & $\pm$ & 59\\
     & 32768 &  1 & 5.0 & Cir & 2833 & 18273.6 & 18.68 & 28.4  & 3120 &       &    & 2848 & & \\
     & 65536 &  1 & 5.0 & Cir & 2833 & 35862.7 & 23.39 & 28.7  & 5092 &       &    & 4635 & & \\
     &131072 &  1 & 5.0 & Cir & 2833 & 71218.1 & 29.40 & 29.2  & 8324 &       &    & 7656 & & \\
\noalign{\smallskip}
(V)  &  8192 &  4 & 5.0 & Cir &15000 &  4489.1 & 35.54 & 31.2  &  7559 & $\pm$ & 126 & 5943 & $\pm$ & 307\\
     & 16384 &  2 & 5.0 & Cir &15000 &  8808.4 & 44.49 & 31.1  & 11495 & $\pm$ & 412 & 8724 & $\pm$ & 641\\
     & 32768 &  1 & 5.0 & Cir &15000 & 18205.4 & 56.67 & 33.1  & 17262 &       &    & 14241 & & \\
     & 65536 &  1 & 5.0 & Cir &15000 & 35914.9 & 71.02 & 32.9  & 26868 &       &    & 22515 & & \\
     &131072 &  1 & 5.0 & Cir &15000 & 71952.3 & 89.59 & 33.7  & 40212 &       &    & 36424 & & \\
\noalign{\smallskip}
(VI) & 32768 &  1 & 5.0 & 0.2 & 8500 & 17981.2 & 29.48 & 30.7  & 7834 &       &    & 6342 & & \\
     & 65536 &  1 & 5.0 & 0.2 & 8500 & 35608.5 & 37.04 & 31.5  &12383 &       &    & 9514 & & \\
\noalign{\smallskip}
(VII)& 32768 &  1 & 5.0 & 0.3 & 8500 & 18300.0 & 25.73 & 31.2  & 6709 &       &    & 5205 & & \\
     & 65536 &  1 & 5.0 & 0.3 & 8500 & 35932.9 & 32.22 & 32.1  &10762 &       &    & 8352 & & \\
\noalign{\smallskip}
(VIII)&32768 &  1 & 5.0 & 0.7 & 8500 & 17957.3 & 12.15 & 28.8  & 3111 &       &    & 2093 & & \\
     & 65536 &  1 & 5.0 & 0.7 & 8500 & 35749.5 & 15.29 & 29.8  & 4940 &       &    & 3470 & & \\
\noalign{\smallskip}
(IX) & 32768 &  1 & 5.0 & 0.8 & 8500 & 18025.7 &  8.94 & 28.1  & 2309 &       &    & 1461 & & \\
     & 65536 &  1 & 5.0 & 0.8 & 8500 & 36116.5 & 11.27 & 29.6  & 3549 &       &    & 2289 & & \\
\end{tabular}
\begin{flushleft}
Notes:  \\

1. For clusters on eccentric orbits, the value given for $R_G$ is the apogalactic distance
\end{flushleft}
\end{table*}

%% file: table2.tex
\begin{table*}
\caption[]{Observed and predicted data for some galactic globular clusters.}
\begin{tabular}{lcrrr@{.}lr@{.}lrrcr}
\noalign{\smallskip}
\multicolumn{1}{c}{Name}& \multicolumn{1}{c}{$M_C$} &
\multicolumn{1}{c}{$R_{P}$} & \multicolumn{1}{c}{$R_{A}$} &  
\multicolumn{2}{c}{$x_{local}$} & \multicolumn{2}{c}{$x_{global}$} &
\multicolumn{1}{c}{$T_{Diss}$} & \multicolumn{1}{c}{$M_0$} & \multicolumn{1}{c}{$F_{CR}$} & 
 \multicolumn{1}{c}{$x_{th}$} \\
 & [$M_\odot$] & [kpc] & [kpc] & \multicolumn{2}{c}{ } & \multicolumn{2}{c}{ } & [GYR] &  [$M_\odot$] & & \\
\noalign{\smallskip}
 NGC 6121 & $1.30 \cdot 10^5$ & 0.6 &  5.9 &   0&0  &  0&0  & 17.3 & $7.5 \cdot 10^5$ &  0.46  &-0.32 \\ 
 NGC 6254 & $1.67 \cdot 10^5$ & 3.4 &  4.9 &  -0&1  & -0&2  & 34.3 & $3.8 \cdot 10^5$ &  0.33  & 0.14 \\ 
 NGC 6341 & $3.26 \cdot 10^5$ & 1.4 &  9.9 &  -0&1  &  0&2  & 35.9 & $7.3 \cdot 10^5$ &  0.32  & 0.16 \\ 
 NGC 6397 & $7.00 \cdot 10^4$ & 3.1 &  6.3 &  -0&5  & -0&35 & 24.3 & $2.2 \cdot 10^5$ &  0.37  & 0.00 \\ 
 NGC 6656 & $4.29 \cdot 10^5$ & 2.9 &  9.3 &  -0&1  & -0&1  & 62.2 & $7.7 \cdot 10^5$ &  0.30  & 0.25 \\ 
 NGC 6752 & $2.11 \cdot 10^5$ & 4.8 &  5.6 &  -0&3  &  \multicolumn{2}{c}{ } &  46.6 & $4.2 \cdot 10^5$ &  0.31  & 0.21 \\
 NGC 6809 & $1.79 \cdot 10^5$ & 1.9 &  5.8 &  -0&35 & -0&35 & 28.6 & $4.7 \cdot 10^5$ &  0.34  & 0.08 \\ 
 NGC 7078 & $7.96 \cdot 10^5$ & 5.4 & 10.3 &   0&1  &  0&4  &135.6 & $1.3 \cdot 10^6$ &  0.29  & 0.29 \\ 
 NGC 7099 & $1.59 \cdot 10^5$ & 3.0 &  6.9 &   0&2  &  0&25 & 34.7 & $3.6 \cdot 10^5$ &  0.32  & 0.15 \\ 
 PAL 5    & $5.5  \cdot 10^3$ & 7.0 & 19.0 &  -0&5  & -0&5  & 17.2 & $3.2 \cdot 10^4$ &  0.46  &-0.32 \\ 
\end{tabular}
\end{table*}